\documentclass[12pt, twocolumn]{article}
\usepackage{lineno} 

\usepackage[left=0.6in, right=0.6in, top=0.8in, bottom=0.8in]{geometry}
\usepackage{graphicx}
\usepackage{amsmath, amssymb}
\usepackage[numbers,sort&compress]{natbib}
\usepackage{titlesec} 
\usepackage{ragged2e}
\usepackage[T1]{fontenc}
\usepackage{placeins}
\usepackage{subcaption}
\usepackage{adjustbox}
\usepackage{booktabs, tabularx, makecell}
\usepackage{multirow}
\usepackage{rotating}
\usepackage{array}
\usepackage{siunitx}
\usepackage[font=small,labelfont=bf]{caption}
\usepackage{enumitem}
\usepackage{breqn}
\usepackage{makecell}
\usepackage{xcolor}
\usepackage{hyperref}

\definecolor{linkblue}{RGB}{0,45,114}
\definecolor{citeblue}{RGB}{0,70,140}
\definecolor{urlblue}{RGB}{0,90,120}
\definecolor{filegray}{RGB}{80,80,80}

\hypersetup{
    colorlinks=true,
    linkcolor=linkblue,
    citecolor=citeblue,
    urlcolor=urlblue,
    filecolor=filegray
}

\newcolumntype{T}{!{\vrule width 1.8pt}}

\titleformat{\section}{\large\bfseries}{\thesection}{1em}{}
\titleformat{\subsection}{\normalsize\bfseries}{\thesubsection}{1em}{}
\newcounter{numquote}

\title{Diversity-Aware Active Learning Builds Materials Datasets for Current and Future Objectives}
\author{
    Rafael Espinosa Castañeda$^{1,4,5}$, Ashley S.~Dale$^{1,4}$, Hongchen Wang$^{1}$, Yonatan Kurniawan$^{1}$, \\
    Hao Wan$^{1}$, Runze Zhang$^{1}$, Adji Bousso Dieng$^{3}$, Kangming Li$^{2}$, Jason Hattrick-Simpers$^{1,5,*}$
}
\date{}
\begin{document}
\nolinenumbers 

\justifying
\twocolumn[
\maketitle
\begin{center}
    {\small 
    $^{1}$ Department of Materials Science and Engineering, University of Toronto, Canada.\\
    $^{2}$ Department of Materials Science and Applied Physics,\\  
    King Abdullah University of Science and Technology, Saudi Arabia.\\ 
    $^{3}$Department of Computer Science, Princeton University \\
    $^{4}$Vector Institute for Artificial Intelligence, Toronto, Canada\\
    $^{5}$Acceleration Consortium, Toronto, Canada\\
    $^{*}$ Corresponding author:jason.hattrick.simpers@utoronto.ca
    }
\end{center}
    
\begin{abstract}
Materials science data collection can be expensive, making dataset reuse and long-term utility important for future discovery. In practice, data collection often prioritizes properties of current interest while other properties are excluded from the acquisition process. This can produce datasets that are well suited for current objectives but lose information needed for properties of future interest. Here, we present how to construct datasets that remain informative for both. We develop and evaluate acquisition frameworks that combine predictive disagreement with diversity in material features, predicted outcomes, or both. We test these approaches across four DFT datasets and an experimental thermoelectric dataset. Diversity inclusion in dataset construction can reduce present and future objectives trade-off by preserving a broader range of materials and property behavior during acquisition. Across all datasets, the Non-dominated Sorting Genetic Algorithm II with disagreements and feature diversity (NSGA-II Dis. + diversity) framework developed in this work provides the most robust balance between current and future objectives. It maintains prediction performance above random sampling for current objectives while mitigating bias on excluded properties and, in many cases, improving their performance beyond random sampling.  This diversity-aware acquisition framework provides a practical route to building materials datasets with unbiased quality entries and mitigating
cold-start limitations in subsequent modeling and discovery campaigns.
\end{abstract}
\vspace{1em}
]
\clearpage

\section{Introduction}
Data that can be reused across different scientific objectives is essential for accelerating discovery while reducing the time and cost of research. In materials science, high-quality experimental and computational data are often expensive to generate, requiring specialized instrumentation~\cite{Liu2026}, complex synthesis~\cite{Solomon2025}, difficult measurements~\cite{pmid15323872}, or large-scale simulations~\cite{10.1177/1094342018819741, doi:10.1142/S0129183106010182, Wines_2024}. Consequently, the long-term value of a dataset depends not only on the quality of the measurements, but also on its ability to support future scientific questions beyond those originally envisioned.

Large community repositories have played a transformative role in enabling such reuse. Databases such as the Materials Project~\cite{TheMaterialsProjectAmaterialsgenome,10.1063/1.4812323}, JARVIS~\cite{JARVISDataBase}, OQMD~\cite{OQMD_DataBase,OQMD_reflections,OQMDSaal2013}, AFLOW~\cite{Divilov2025,OSES2023111889}, and ICSD~\cite{Zagorac:in5024,Allmann:sh0188,Belsky:an0615} have supported a broad range of data-driven materials discovery workflows. These repositories have enabled materials characterization~\cite{Lee2020DeepLearningXRD,10.1093/nsr/nwaf421}, predictive modeling of materials properties~\cite{PhysRevLett.120.145301, choudhary_atomistic_2021,goodall_predicting_2020, jha_elemnet_2018,isayev_universal_2017}, generative and inverse-design models based on crystal structures~\cite{Dan2020, NOH20191370,Xiao2023,Zeni2025}, and a wide spectrum of computational and experimental discovery pipelines~\cite{Ward2016,Kusne2020,Szymanski2023,Pogue2023,Palizhati2022}. By providing shared resources for the scientific community, these datasets allow existing measurements to be repurposed for new research questions and help mitigate the cold-start problem faced by emerging applications with limited task-specific data.

Despite their success, comparatively little attention has been given to how datasets mantain their usefulness beyond the intended initial objectives of their construction. In practice, many data acquisition campaigns are guided by specific research objectives and therefore prioritize a limited set of target properties. For instance, many active learning and Bayesian optimization workflows in materials discovery focus on optimizing a single property at a time~\cite{Kusne2020,XUE2017532,Xue2016,doi:10.1073/pnas.1607412113,wang2025trainingfreeactivelearningframework,Xin2021}. Extensions to bi-objective optimization and involving three~\cite{Rebuffi:23} outcomes have also been implemented. As more outcomes are included, these implementations become more uncommon in materials science. Although such strategies are effective for achieving immediate objectives, they implicitly concentrate sampling in narrow regions of materials space that are most informative for the targeted properties. As a consequence, datasets constructed in this manner may be poorly suited for future modeling tasks or broader scientific reuse.

On the other hand, many material science studies have focused on how to collect data to improve model generalization for specific targets. 
For example, Smith et al. \cite{10.1063/1.5023802} showed how to construct datasets for universal machine learning potentials. Hastings et al. \cite{10.1039/d5dd00361j} investigated how initialization strategies influence Bayesian optimization trajectories in materials discovery campaigns. Bernstein et al. \cite{bernstein2019de} constructed structurally diverse datasets to train transferable interatomic potentials across different configurations and crystal phases. Together, these studies have advanced our understanding of how to build datasets that support out-of-distribution prediction for specific targeted properties from the construction of the dataset. 

Also, many diversity-aware Bayesian optimization methods focus on selecting diverse solutions within high-performing candidates\cite{eriksson2020scalableglobaloptimizationlocal,maus2023discoveringdiversesolutionsbayesian, 10.1145/2783258.2783360,nguyen2024qualityweightedvendiscoresapplication,pmlr-v139-malkomes21a,nava2022diversifiedsamplingbatchedbayesian}. Moreover, different Batch Bayesian Optimization methods have been developed~\cite{pmlr-v97-gong19b,pmlr-v151-nava22a,10.5555/3157382.3157568} and applied in materials science~\cite{Shibukawa2026,Couperthwaite2020,WILSON2022111330,HASTINGS2025121173,Alvi2026} where samples to be selected for the design targets are spread in the chemical space. These strategies are valuable for targeted discovery, but they remain tied to the objectives used during acquisition. Nevertheless, it is unclear if current state of the art methods for dataset construction for optimizing specific material properties also preserve predictive performance for properties excluded from the selection objective.  To the best of our knowledge, no attention has been given to building materials datasets that maintain strong predictive performance on the selected targets while mitigating bias for properties that were not considered during data collection.

Furthermore, recent evidence further suggests that simply increasing dataset size does not guarantee improved scientific utility. Li et al.~\cite{RedundancyMaterialsData} showed that widely used repositories such as Materials Project, JARVIS and OQMD contain substantial redundancy. For single-objective tasks, $70$--$95\%$ of the training data can be removed while keeping the degradation in test-set predictive performance below $10\%$ on newer database versions. These findings indicate that many datasets oversample redundant regions of materials space while underrepresenting the broader diversity of materials and properties. As a result, data acquisition strategies that focus solely on expanding dataset size may fail to improve the informativeness or general utility of the resulting datasets.

These observations raise a fundamental question: how should materials datasets be constructed so that they remain informative not only for immediate research objectives, but also for future scientific tasks that were not anticipated at the time of data collection?

Here, we present and evaluate seven frameworks for materials dataset construction. The frameworks are based on sampling strategies that have been widely used in materials science. Two serve as focused baselines: query by committee (QBC) and hypervolume-based acquisition (HV). We develop five additional frameworks that combine model disagreement with different forms of diversity. One promotes diversity in feature space, two promote diversity in predicted outcomes, and two combine feature- and outcome-space diversity. This design allows us to separate the effect of informative sampling from the effect of diversity and to determine which combinations best preserve information for both current and future objectives.

\begin{figure*}[!t]
    \centering
\includegraphics[width=0.95\textwidth]{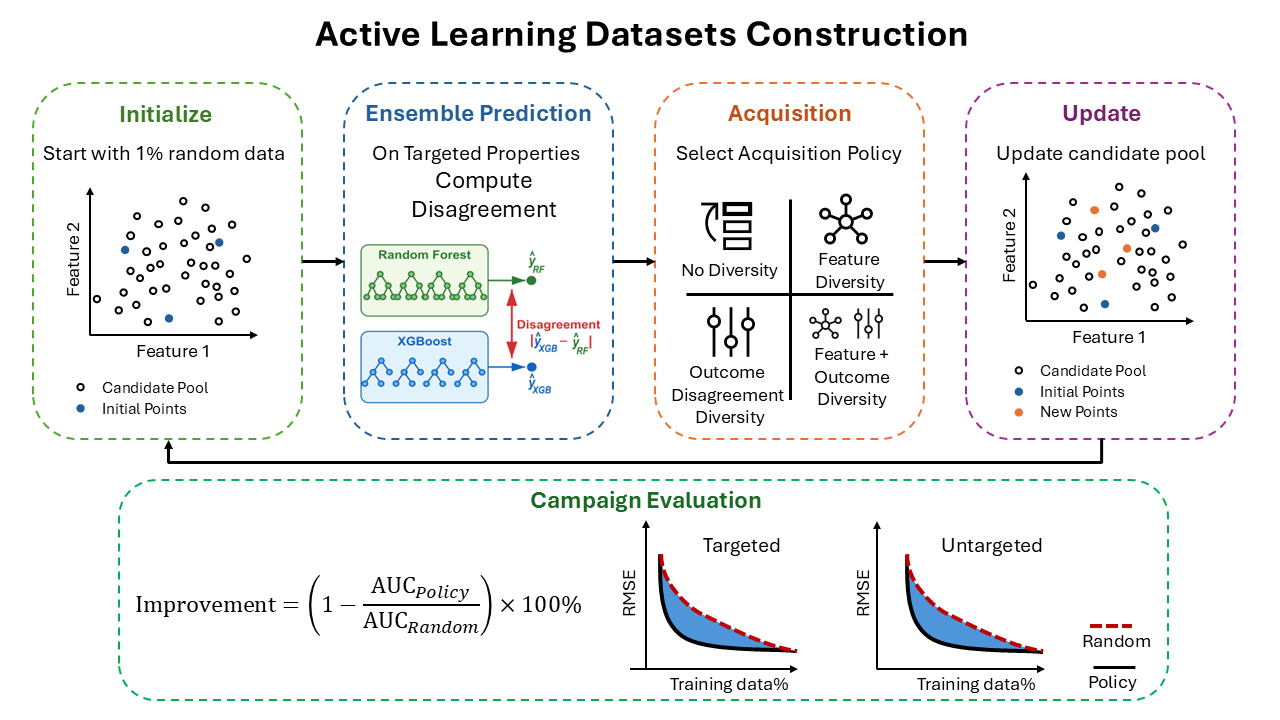}
\caption{\textbf{Schematic of the active-learning workflow used for dataset construction.} Each curated dataset is first divided into a fixed $10\%$ test set and a $90\%$ acquisition set. The campaign starts from $1\%$ randomly selected training data, while the remaining samples form the candidate pool. At each iteration, Random Forest and XGBoost models are trained independently for each targeted property, and candidate informativeness is estimated from the min--max normalized disagreement between their predictions. Acquisition policies then select a batch equal to $1\%$ of the original acquisition dataset using different combinations of disagreement, outcome-space diversity, and feature-space diversity. Only targeted properties contribute to the acquisition decision; untargeted properties are excluded and used to evaluate how informative the resulting dataset remains for future objectives. After each acquisition step, the selected samples are added to the training set and removed from the candidate pool. Predictive performance is quantified by improvement, defined as the percentage reduction in the area under the RMSE learning curve (RMSE-AUC) relative to random sampling on the fixed test set. The blue shaded area represents $\mathrm{AUC}_{\mathrm{Random}}-\mathrm{AUC}_{\mathrm{Policy}}$. Improvement is calculated by dividing this difference by $\mathrm{AUC}_{\mathrm{Random}}$ and multiplying by $100\%$. Positive values indicate better predictive performance over the campaign.
}
\label{fig:DivActiveLearningLoop}
\end{figure*}

To evaluate these frameworks, we design a pool-based active learning workflow. Acquisition policies are assessed according to the performance of surrogate models trained on the selected samples. The accuracy of the models is measured on both properties considered during the dataset construction and those properties which are not of current interest. We call along the paper these properties as targeted (properties of current interest in the dataset construction) and untargeted (future objective properties). The analysis is performed under single, double, and triple-objective acquisition settings in order to evaluate how increasing the number of properties influences dataset quality.

We test these strategies across five datasets comprising both computational and experimental materials data: four density functional theory (DFT) repositories and one experimental thermoelectric dataset composed. Across these settings, we show that the framework NSGA-II Dis. + diversity is a robust sampling strategy that constructs datasets that remain highly informative for targeted properties while preserving predictive performance for outcomes that were not considered during dataset construction.

\section{Results and Discussion}
\label{results_discussion}
We evaluated the dataset construction frameworks using the active-learning workflow shown in Fig. \ref{fig:DivActiveLearningLoop}. The workflow includes dataset initialization, model training, candidate selection, acquisition, and performance evaluation against random sampling. Before active learning, we randomly assigned $10\%$ of each curated dataset to a fixed, independent test set $T'$. The specific information of the curated datasets can be found in Section \ref{subsec1}. The remaining $90\%$ formed the acquisition dataset. From this acquisition dataset, $1\%$ of the samples were randomly selected as the initial training set $T_0$, and the remaining samples formed the initial candidate pool $\mathcal{U}_0$. At each acquisition iteration $t$, a policy selected a batch $B_t \subseteq \mathcal{U}_t$, corresponding to $1\%$ of the original acquisition dataset. The selected samples were then transferred from the candidate pool to the training set:
\[
T_{t+1}=T_t\cup B_t,
\qquad
\mathcal{U}_{t+1}=\mathcal{U}_t\setminus B_t.\]
To assess sensitivity to the initial training data, the complete acquisition campaign was repeated using five random seeds.
At each iteration, we trained one Random Forest (RF) regressor and one XGBoost (XGB) regressor for each target property using the current training set $T_t$. All models were single-output regressors; therefore, dependencies among target properties were not imposed at the model level. RF and XGB were selected because they provide strong predictive performance on tabular data at a computational cost compatible with repeated model retraining during active learning~\cite{10.5555/3600270.3600307}. Model hyperparameters are reported in the Supplementary Information section ~\ref{model_hyperparameters}.

The trained models were evaluated on the fixed test set $T'$ at every iteration. We recorded the coefficient of determination $R^2$, root-mean-square error (RMSE), and mean absolute error (MAE) for each property. 

As a measure of informativeness, we used the min–max-normalized absolute disagreement between the XGB and RF surrogate models for the targeted outcomes. All acquisition policies incorporate this informativeness measure when selecting new samples. In contrast, the untargeted outcomes are not used by the acquisition policies and are treated as potential future objectives, allowing us to evaluate how well the resulting datasets support learning tasks that were not explicitly considered during data collection.
\subsection{Acquisition policies}
\begin{figure*}[!t]
    \centering
\includegraphics[
    width=.8\textwidth
]{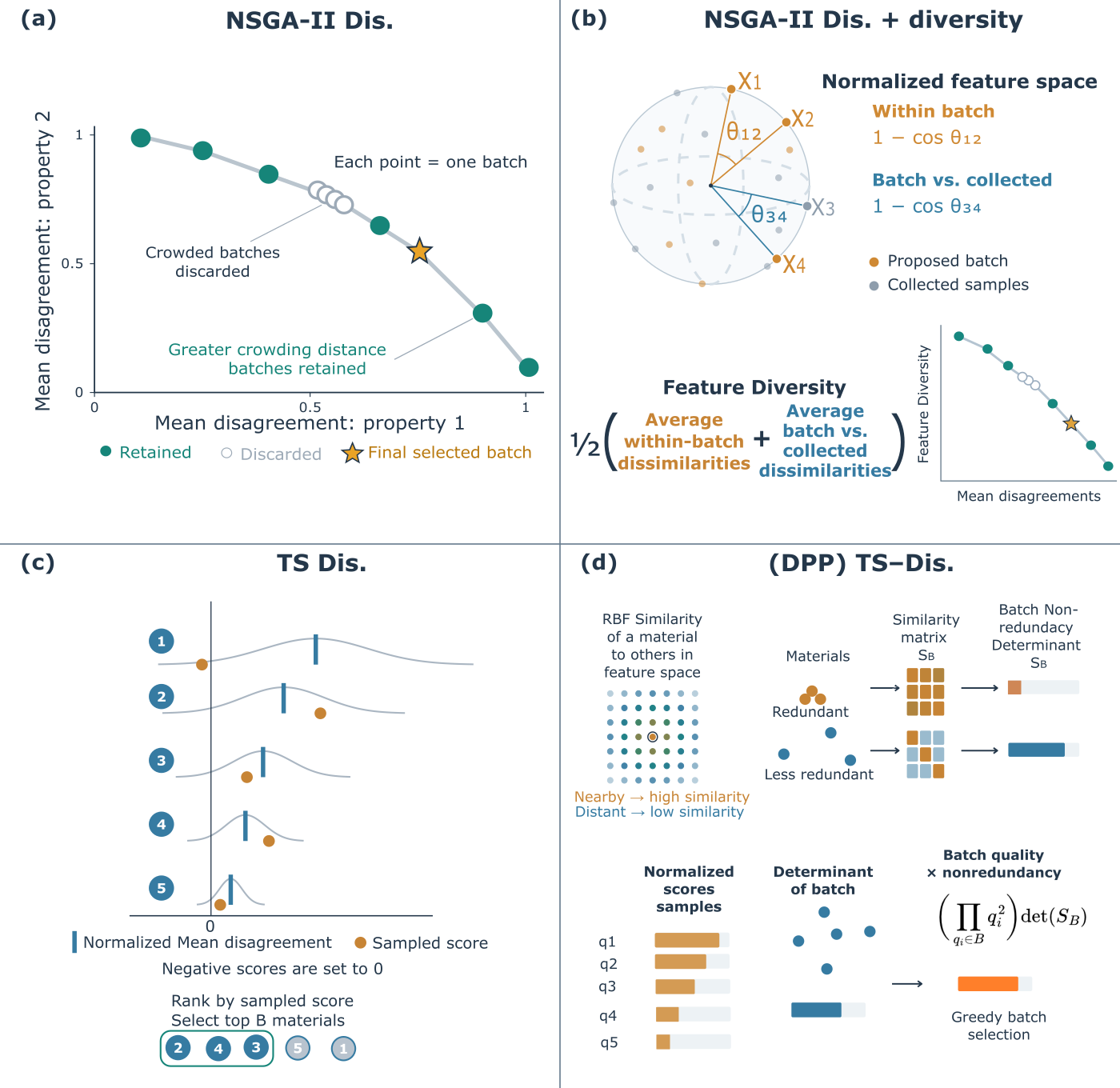}
\caption{\textbf{Disagreement-based batch selection policies.}
(a) NSGA-II compares crowding distances between batch candidates. Larger crowding distances are preferred, promoting diversity across batch-mean disagreement values. One batch is then selected from the found Pareto front by maximizing the equally weighted mean of its front-normalized objectives. (b) NSGA-II Dis. + diversity additionally optimizes feature dissimilarity within a proposed batch and relative to collected samples, with equal weights. (c) TS Dis. draws a Gaussian score for each material, with mean and spread determined by its average normalized disagreement, clips negative scores to zero, and selects the top $B$ (d) DPP TS--Dis. combines normalized sampled scores $q_i$ with RBF feature similarities $S_{ij}$ through $L_{ij}=q_iS_{ij}q_j$. Greedy batch construction approximately maximizes
$\det(L_B)=\left(\prod_{i\in B}q_i^2\right)\det(S_B)$,
balancing score quality and feature nonredundancy. Illustrations are schematic.}
\label{fig:batch_selection_policies}
\label{fig:policies}
\end{figure*}
We compare dataset construction frameworks that differ in how they use models disagreement and diversity. Full mathematical definitions and optimization procedures for all acquisition policies are provided in Section \ref{sec:acquisition_policies}. 

The acquisition policies select batches using normalized predictive
disagreement between RF and XGB models for each targeted property. Query by Committee (QBC)
sums these disagreements, whereas hypervolume disagreement (HV)
uses their product to favor candidates with high disagreement across
all targets. The Non-dominated Sorting Genetic Algorithm II with
disagreement objectives (NSGA-II Dis.) optimizes mean normalized
disagreement for each target as a separate batch objective.
It maintains candidate batches with different target-specific
disagreement trade-offs. When a nondominated front exceeds the
remaining population capacity, batches with larger crowding
distances are preferred (Fig.~\ref{fig:batch_selection_policies}a). The final batch maximizes the equally
weighted mean of objectives normalized over the resulting Pareto set.

Feature diversity measures dissimilarity between material descriptors.
The Non-dominated Sorting Genetic Algorithm II with disagreement
and feature-diversity objectives (NSGA-II Dis. + diversity) adds an
objective that equally weights average cosine dissimilarity within
the proposed batch and between the batch and collected samples.
This measure uses unit-normalized feature vectors (see Fig.~\ref{fig:batch_selection_policies}b).
Hypervolume disagreement with $k$-medoids clustering
(HV + $k$-medoids) clusters the top $10\%$ of candidates ranked
by HV in feature space and selects one representative medoid
per cluster.

Thompson Sampling using model disagreements (TS-Dis.) introduces
stochastic exploration through a Thompson-inspired score
perturbation. It independently perturbs each candidate's mean
disagreement with Gaussian noise. Negative scores are clipped to
zero, and the candidates with the highest scores are selected (Fig.~\ref{fig:batch_selection_policies}c).
This stochastic ranking allows candidates with lower mean
disagreement to be selected in a proposed batch and can broaden the selected
disagreement values. Determinantal Point Process Thompson Sampling
using model disagreements (DPP TS-Dis.) retains the top $10\%$
of these scored candidates. It uses greedy determinant maximization
to balance score-based quality with low feature redundancy,
measured through a radial basis function (RBF) feature-similarity
kernel (Fig.~\ref{fig:batch_selection_policies}d).

NSGA-II Dis., NSGA-II Dis. + diversity, TS-Dis., and DPP TS-Dis.
can promote outcome-disagreement diversity through
different selection mechanisms. NSGA-II maintains candidate batches with
different disagreement trade-offs and favors less crowded Pareto
solutions when a front must be truncated.
TS-based policies stochastically alter candidate rankings,
allowing samples with lower mean disagreement to be selected. However, neither mechanism guarantees outcome-disagreement
diversity within the final batch. On the other hand, feature diversity is explicitly
incorporated through a dissimilarity objective in
NSGA-II Dis. + diversity, determinant-based selection in
DPP TS-Dis., and feature-space clustering in HV + $k$-medoids.

Table~\ref{tab:policy_diversity_table} groups the policies by their
diversity mechanisms. Acquisition uses only targeted properties;
performance on untargeted properties is evaluated separately.

\begin{table}[htbp]
    \centering
    \small
    \setlength{\tabcolsep}{3pt}
    \renewcommand{\arraystretch}{1.15}

    \begin{tabularx}{\columnwidth}{
        @{}*{3}{>{\raggedright\arraybackslash}X}@{}
    }
        \hline
        \textbf{Outcome-disagreement only} &
        \textbf{Feature only} &
        \textbf{Both} \\
        \hline
        NSGA-II Dis. &
        HV + $k$-medoids &
        NSGA-II Dis. + diversity \\
        TS-Dis. & & DPP TS-Dis. \\
        \hline
    \end{tabularx}

    \caption{Acquisition policies grouped by diversity mechanism.
    Outcome-disagreement diversity can be promoted but is not
    guaranteed within the selected batch. Feature diversity is
    explicitly included in the optimization or selection criterion.}
    \label{tab:policy_diversity_table}
\end{table}

Across all datasets, we do not notice significative differences between QBC and hypervolume sampling. So, we omit the results of the last policy for all the main text and its results can be found in the Supplementary Information.
\begin{figure*}[t]
    \centering
\includegraphics[width=1.\textwidth]{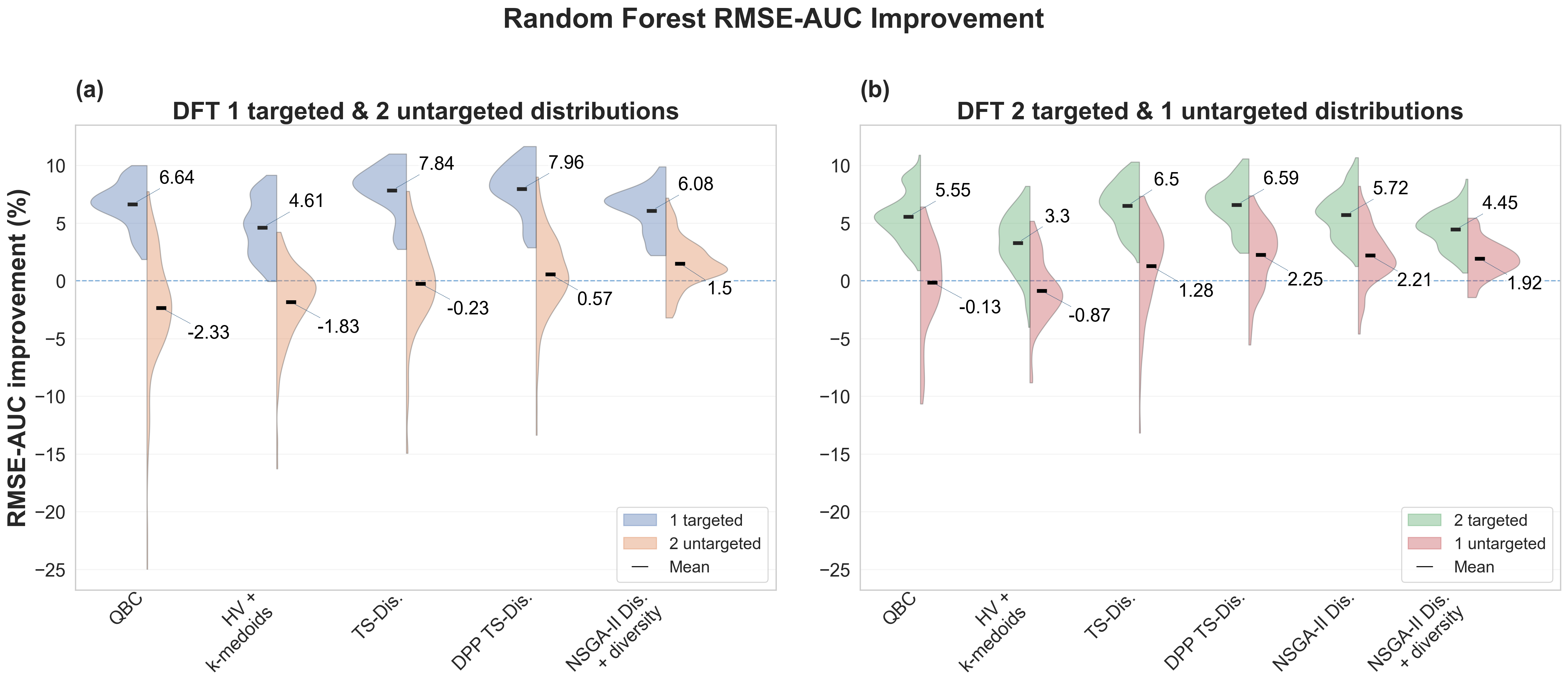}
\caption{\textbf{Distributions of RMSE-AUC improvement for one- and two-property acquisition cases across all DFT datasets using Random Forest.} 
}
\label{fig:DFT_ImprovementAll}
\end{figure*}
\subsection{DFT dataset construction}
We evaluated active-learning dataset construction across four DFT datasets, MP18, MP21, JARVIS18, and JARVIS22, using band gap, bulk modulus, and formation energy as potential prediction targets. Campaigns considered either one or two acquisition objectives, while the remaining two or one properties, respectively, were treated as untargeted. Predictive performance was measured by the percentage reduction
in the area under the RMSE learning curve (RMSE-AUC) relative
to random sampling, defined here as improvement. Figure~\ref{fig:DivActiveLearningLoop} illustrates this calculation. 

The results include five random seeds and all corresponding target-property combinations across the four datasets. Pairwise Pearson correlations between the three properties are lower or equal to 0.61 in magnitude. Thus, improved prediction of an untargeted property cannot be explained simply by a strong linear correlation with an acquisition target. Explicit performance formulation can be found in section \ref{subsec_performance_measure} and individual correlations are reported in the Supplementary Information section \ref{Corr_matrices}.

Acquisition focused only on targeted properties improves their prediction. However, this gain can come at the expense of properties excluded from the acquisition objective. This trade-off is evident for one-target QBC in Fig.~\ref{fig:DFT_ImprovementAll}(a), which shows the Random Forest results. Across the four DFT datasets, QBC achieves a mean RMSE-AUC
improvement of $6.64\%$ for the targeted property with Random Forest. In contrast, for the two untargeted properties, the mean improvement is
$-2.33\%$. The same imbalance is observed for MP18, MP21, JARVIS18, and JARVIS22 and when using XGBoost as regressor. These other results are reported in the Supplementary Information section \ref{improvement_dist}.

Increasing the number of targeted properties mitigates this bias, but does not eliminate it. This is shown in Fig.~\ref{fig:DFT_ImprovementAll}(b). With two targeted properties, QBC achieves mean targeted RMSE-AUC improvement of $5.55\%$ and untargeted still below random sampling with Random Forest.

In general, diversity reduces this effect. As shown in Figs. ~\ref{fig:DFT_ImprovementAll}(a,b), the distributions for untargeted properties shift toward random sampling or above it. This trend holds for both one and two untargeted properties. Nevertheless, not all diversity dataset construction policies mitigate bias equally.

Model disagreement diversity policies substantially mitigates bias for future objectives. For example, in one-target campaigns, TS-Dis. achieves targeted RMSE-AUC improvements of $7.84\%$ with Random Forest while the mean improvement in the two untargeted properties is shifted to be only $0.23\%$ below random sampling. With two targeted properties, the remaining untargeted property improves above random sampling by $1.28\%$. The same trend can be observed with NSGA-II Dis. in Fig. ~\ref{fig:DFT_ImprovementAll}(a,b). 

Feature-space diversity alone does not provide the same benefit. The untargeted properties do not improve consistently. When targeting one property, HV + $k$ medoids outperformed focused QBC on untargeted property prediction Fig. ~\ref{fig:DFT_ImprovementAll}(a). However, when targeting two properties, its performance on the remaining untargeted property fell below that of focused QBC Fig. ~\ref{fig:DFT_ImprovementAll}(b). Similar trends are observed (Fig. ~\ref{fig:DFT_ImprovementAll}(b)) when predicting with XGBoost, which can be found in the Supplementary Information section \ref{improvement_dist}.

\begin{figure*}[t]
    \centering
\includegraphics[width=.8\textwidth]{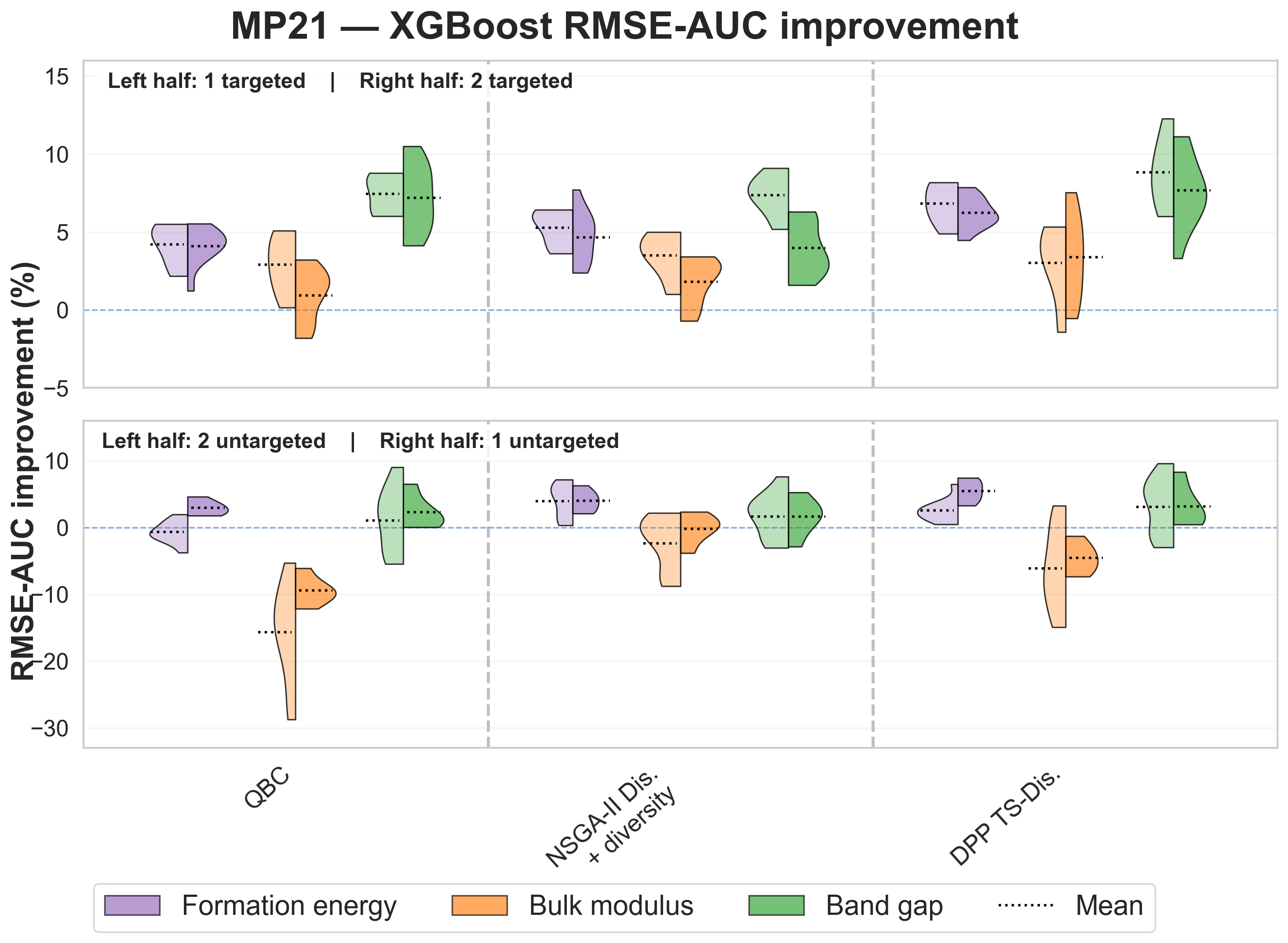}
\caption{\textbf{XGBoost RMSE-AUC improvement distributions for formation energy, bulk modulus, and band gap in MP21.} This dataset is shown because its larger variability across properties and acquisition cases provides a robust test of policy behavior. In this higher-variance dataset, NSGA-II Dis. + diversity provides the most balanced overall performance across targeted and untargeted properties.
}
\label{fig:MP21_improvement1_2}
\end{figure*}

The strongest overall balance is obtained when both disagreement and feature-space diversity are considered. In one target property setting, DPP TS-Dis. shows the highest mean improvement for targeted property, whereas NSGA-II Dis. + diversity performs better on average for the untargeted properties. When extending to two targets, DPP TS-Dis. shows the best performance on targeted and untargeted properties. However, the distributions for DPP TS-Dis. extend further below zero and include more negative values than those for NSGA-II Dis. + diversity (Figs.~\ref{fig:DFT_ImprovementAll}(a,b)). This indicates more frequent and greater underperformance relative to random sampling. When two properties are untargeted, DPP-TS-Dis. falls below random sampling in $50$ of $120$ cases, compared with $20$ of $120$ cases for NSGA-II Dis. + diversity. When one property is untargeted, these numbers correspond to $14$ and $6$ of $60$ cases, respectively. Also, with two untargeted properties, DPP-TS-Dis. largest decrease relative to random sampling is $13.36\%$, compared with $3.20\%$ for NSGA-II Dis. + diversity. With one untargeted property, the corresponding decreases are $5.54\%$ and $1.44\%$. Across all policies evaluated, NSGA-II Dis. + diversity had the fewest cases of underperformance relative to random sampling and the smallest maximum decrease in improvement relative to random sampling. These results strongly suggest that NSGA-II Dis. + diversity provides the most robust mitigation of bias for untargeted properties.

\begin{figure*}[t]
    \centering
\includegraphics[width=.89\textwidth]{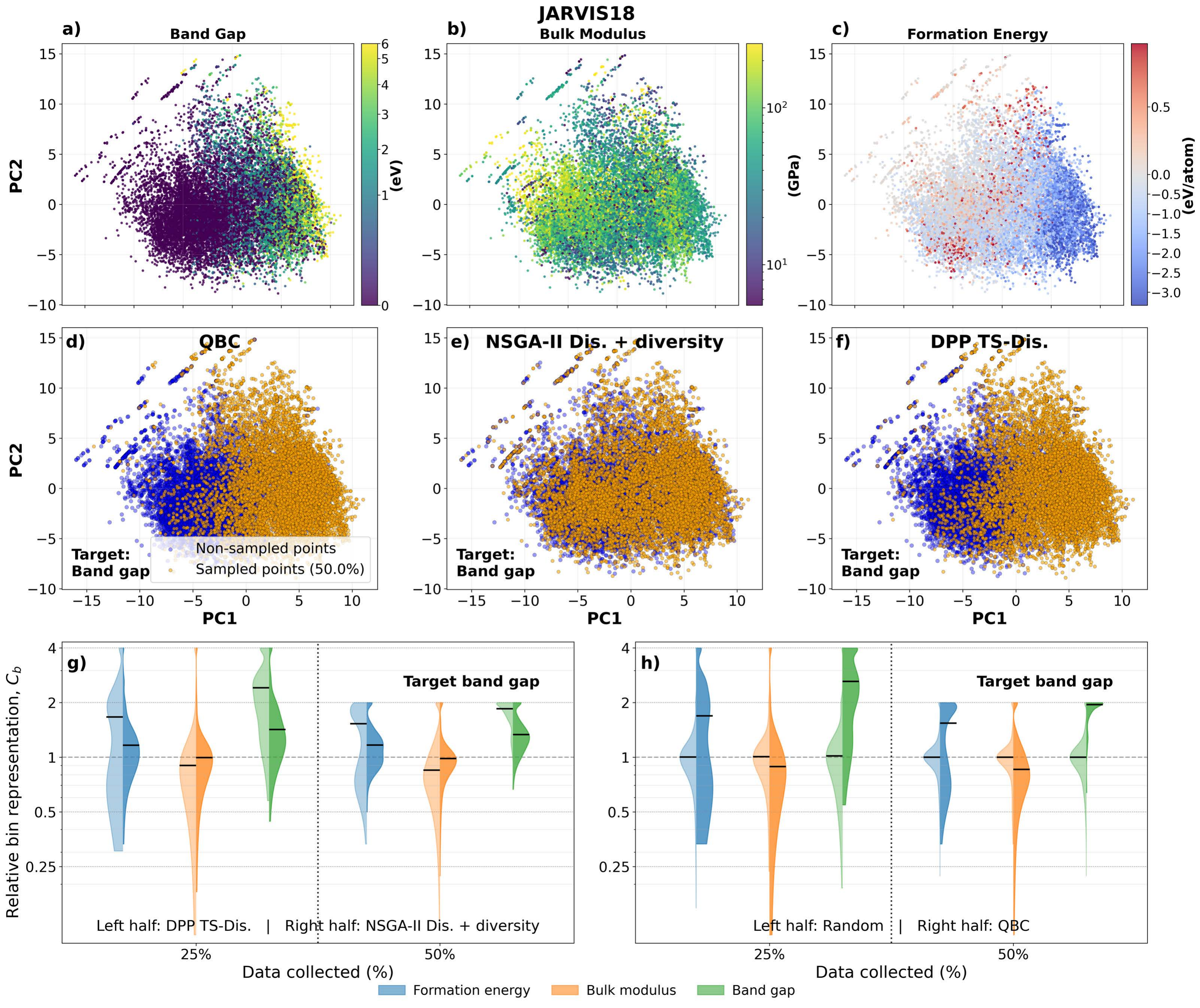}
\caption{\textbf{Sampling in JARVIS18 when targeting band gap.} \textbf{a--c}, First two principal components, colored by band gap, bulk modulus, and formation energy. \textbf{d--f}, Samples selected by QBC, NSGA-II Dis. + diversity, and DPP-TS-Dis. after collecting 50\% of the acquisition pool. \textbf{g,h}, Relative sampling distributions (Eq.~\ref{relative_sampling}); values above, below, and equal to 1 indicate over-, under-, and proportional sampling relative to the complete pool.
}
\label{fig:jarvis18_pca_violin}
\end{figure*}

Analysis at the property level shows that overall trends hold across different targeted and untargeted properties. Figure~\ref{fig:MP21_improvement1_2} compares QBC and the two best-performing policies on MP21. Among the DFT datasets, MP21 showed the greatest variation in RMSE-AUC improvement across properties. This case helps assess whether each policy performs consistently when the targeted properties change. The other cases show less variation and the same overall trends, as reported in Supplementary Information section \ref{improvement_dist}. When left untargeted, bulk modulus shows the greatest degradation in predictive performance relative to random sampling. NSGA-II Dis. + diversity achieves the best average performance for untargeted bulk modulus, at the expense of some performance on the targeted properties. Nevertheless, performance on the targeted properties remains well above random sampling. Furthermore, its performance is comparable to DPP TS-Dis. for the other two properties when untargeted. 

The origin of the difficulty of bulk modulus as untargeted property can be understood from the structure of the feature and outcome spaces shown in Fig.~\ref{fig:jarvis18_pca_violin}. Panels (a--c) show the three outcomes projected onto the same two principal components of feature space. Band gap and formation energy exhibit visible organization. In contrast, bulk-modulus values are strongly intermixed across neighboring regions in panel (b). Thus, materials that are close in this feature representation can have substantially different bulk moduli. This weaker local organization is consistent with the greater prediction difficulty observed for bulk modulus.  

Focused QBC acquisition and DPP TS-Dis. amplify this difficulty. For example, when band gap is targeted, QBC concentrates the selected materials in specific regions of feature space (Fig.~\ref{fig:jarvis18_pca_violin}(d)). These regions are informative for the current target, but they need not preserve the distributions of bulk modulus or formation energy.  DPP TS-Dis. uses feature and outcome diversity but it still concentrates the sampling in certain regions of the feature space (Fig.~\ref{fig:jarvis18_pca_violin}(f)). In contrast, NSGA-II Dis. + diversity spans broadly the chemical space as shown in figure Fig.~\ref{fig:jarvis18_pca_violin}(e), covering informative cases for bulk modulus. 

To quantify this effect, we divided the range of each outcome into $\sqrt{N_{\mathrm{pool}}}$ bins and measured how frequently each bin was sampled relative to the complete pool. We define the relative sampling of outcome $i$ in bin $b$ at iteration $t$ as

\begin{equation}
C_{i,b,t}=
\frac{N_{i,b,t}/N_t}
{N_{\mathrm{pool},i,b}/N_{\mathrm{pool}}},
\label{relative_sampling}
\end{equation}

where $N_{i,b,t}$ is the number of selected samples whose value for outcome $i$ falls within bin $b$, and $N_t$ is the total number of samples selected at iteration $t$. Similarly, $N_{\mathrm{pool},i,b}$ and $N_{\mathrm{pool}}$ are the corresponding counts for the complete pool. Thus, $C_{i,b,t}=1$ indicates that bin $b$ is sampled at the same relative frequency as in the complete pool for outcome $i$. Values above or below one indicate over- or under-sampling, respectively. 

The split violins in Fig.~\ref{fig:jarvis18_pca_violin}(g,h) show the relative sampling of different outcome ranges during dataset construction when targeting band gap. DPP-TS-Dis. tends to oversample specific ranges of band gap and formation energy, while several bulk-modulus ranges remain undersampled. A similar behavior is observed for QBC. In both cases, the data construction  distortion the distribution of the untargeted outcomes. Random sampling, as expected, remains centered around $C_b=1$, indicating that the sampled data approximately preserve the full outcome distributions.

NSGA-II Dis. + diversity shows a different behavior. The targeted band-gap distribution remains biased toward specific ranges, consistent with preferentially selecting informative samples for the target. In contrast, the untargeted properties median shift closer to $C_b=1$ during acquisition. By treating diversity as an additional objective, NSGA-II Dis. + diversity therefore limits the bias toward specific ranges of the untargeted distributions. This behavior is consistent with its greater robustness for untargeted properties. The method reduces some of the improvement obtained for the current target, while remaining above random sampling, but better preserves the overall distributions of untargeted properties during dataset construction.

Taken together, the DFT results show a clear trend. Focused acquisition can highly reduce performance on excluded properties. Promoting diversity in model disagreement substantially mitigates this bias, and combining it with feature-space diversity provides the best overall balance. Feature-space diversity alone is insufficient. Overall, NSGA-II Dis. + diversity provides the strongest  balance framework for targeted and untargeted properties. It shows the most robust dataset construction framework, with less extreme long distribution negative improvement  cases and it can mitigate bias even for difficult prediction cases. Importantly, these trends are reproduced with both Random Forest and XGBoost. Their consistency across two distinct surrogate models strongly suggests that the observed behavior is driven by the information retained in the acquired datasets, rather than by a model-specific response to the acquisition policies. The model results not shown in this section and the dataset-level results are provided in the Supplementary Information section \ref{improvement_dist}.

\subsection{Experimental Datasets construction}
We evaluated active-learning dataset construction using the sysTEM dataset as the candidate pool and four thermoelectric properties: electrical conductivity, thermal conductivity, Seebeck coefficient, and figure of merit $zT$. One, two, or three properties were used as acquisition objectives, while the remaining properties were treated as untargeted. As for the DFT datasets, predictive performance was quantified by the percentage reduction in the area under the RMSE learning curve (RMSE-AUC) relative to random sampling, defined here as improvement. Figure~\ref{fig:DivActiveLearningLoop} illustrates this calculation. Pairwise Pearson correlations between the four properties were at most $0.45$ in magnitude. Therefore, improved prediction of an untargeted property cannot be attributed simply to strong linear correlation with an acquisition target. Individual correlations are reported in the Supplementary Information section \ref{Corr_matrices}.
\begin{figure*}[t]
    \centering
\includegraphics[width=1.\textwidth]{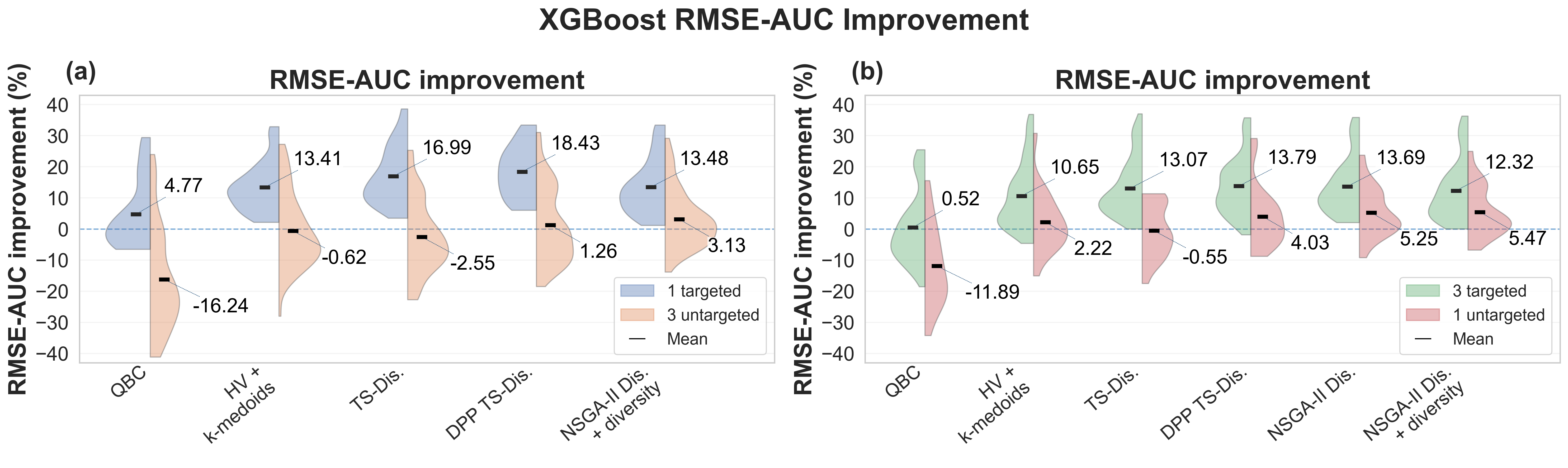}
\caption{\textbf{(a-b) RMSE-AUC improvement distributions for the sysTEM experimental dataset with three targeted properties and one untargeted property across all target combinations using XGBoost.} Positive values indicate better performance than random sampling.
}
\label{fig:Exp_ImprovementAll}
\end{figure*}

The experimental results show a stronger benefit from diversity than the DFT results. For the DFT datasets, focused QBC acquisition generally improved targeted properties at the expense of untargeted ones. For the experimental dataset, focused acquisition can also provide only small gains for the properties it directly targets. This distinction is evident for the one and three-target XGBoost cases in Figs.~\ref{fig:Exp_ImprovementAll}(a,b). On average, QBC improves targeted RMSE-AUC  over random sampling by only $4.77\%$ and $0.52\%$, respectively. In contrast, diversity policies targeted mean RMSE-AUC improvements are above $10\%$. Thus, restricting acquisition to disagreement alone does not maximize performance even for the objectives guiding data collection.

Additionally, diversity policies shift the RMSE-AUC improvement distribution of untargeted properties toward random sampling or above it, with higher mean improvements for untargeted properties than non-diverse counterpart (Fig.~\ref{fig:Exp_ImprovementAll}(a,b)). With three untargeted properties, NSGA-II Dis. + diversity and DPP TS-Dis. perform above random sampling, while the other diversity policies remain slightly below it (Fig.~\ref{fig:Exp_ImprovementAll}(a)). On average with one untargeted property, nearly all diversity policies perform above random sampling, except TS-Dis. (Fig.~\ref{fig:Exp_ImprovementAll}(b)).

In general, the experimental results show the same ranking when changing the number of  untargeted properties across the different  policies. Nevertheless, interestingly, when only one target is left out, the performance of this with HV + $k$ medoids becomes greater than random sampling. Furthermore, its performance for the targeted properties are comparably to the other diversity methods and behaving better than TS-Dis. for untargeted property, which in this last case on average still is below random sampling. In contrast to DFT datasets, in this experimental case, feature diversity alone when increasing the number of properties can have strong performance on future objectives while maintaining good performance on the targeted property.

\begin{figure*}[t]
    \centering
\includegraphics[width=.8\textwidth]{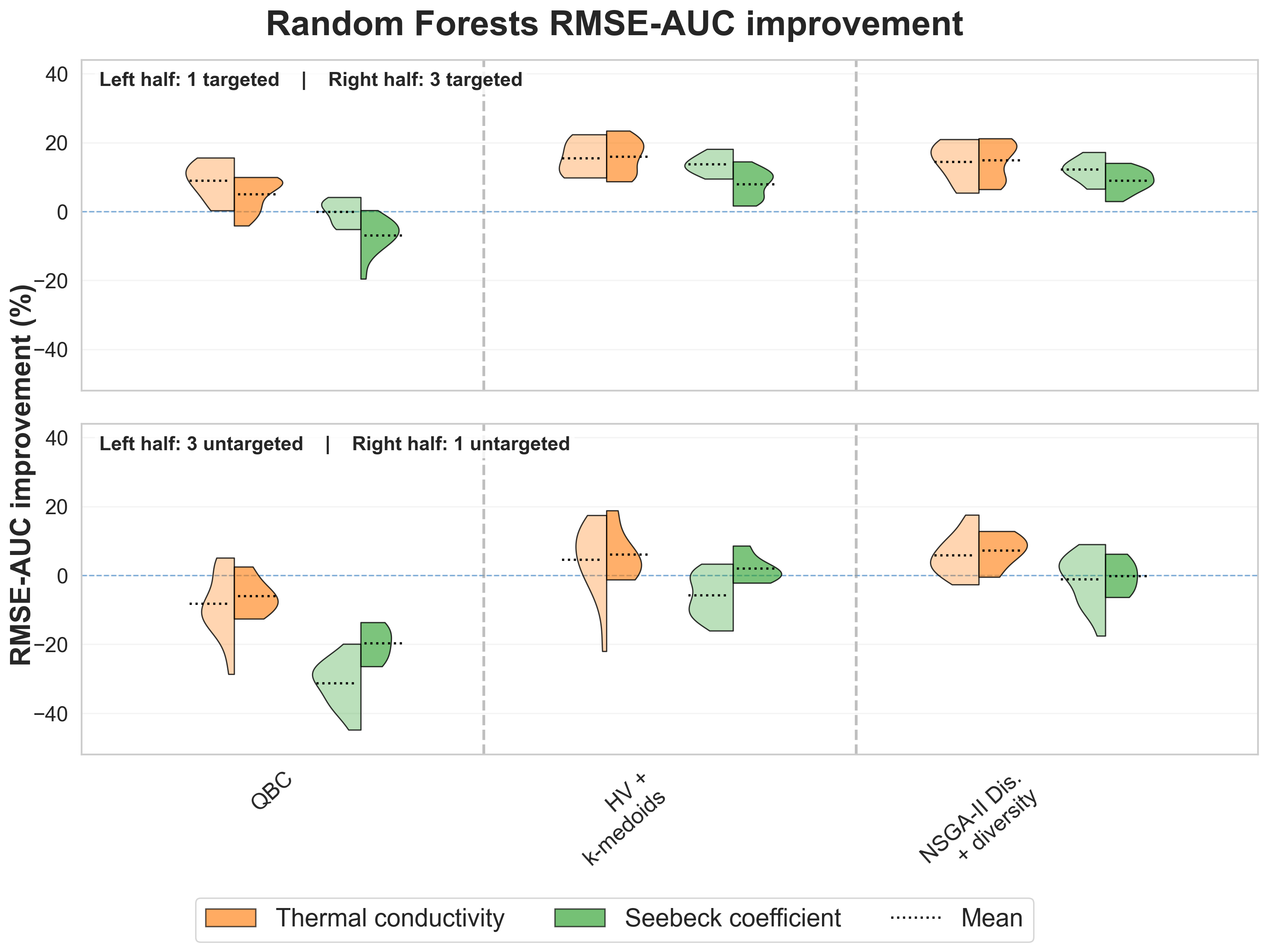}
\caption{\textbf{Property-level RMSE-AUC improvement distributions for the sysTEM experimental dataset using Random Forest.} The upper panel compares each property when targeted alone and when included among three acquisition targets, while the lower panel compares the same properties when three or one objectives are left untargeted. Diversity-based policies consistently shift the distributions toward random sampling or better across all four properties. Nevertheless, the most robust policy is found with NSGA-II Dis. + diversity.
}
\label{fig:threecases_onecase_exp}
\end{figure*}
Property-level analysis shows that the benefit of diversity is consistent across the experimental outcomes, although its magnitude depends on the property. Seebeck coefficient and $zT$ show the lowest RMSE-AUC improvements across policies. For clarity, Fig.~\ref{fig:threecases_onecase_exp} shows two properties: Seebeck coefficient, with among the lowest RMSE-AUC improvements, and thermal conductivity, with among the highest. It compares QBC with NSGA-II Dis. + diversity, the strongest policy overall, and HV + $k$-medoids, which performs competitively on this dataset despite its weaker performance on the DFT datasets. Results for the remaining properties and policies are provided in Supplementary Information Section~\ref{improvement_dist}. For Seebeck coefficient, QBC performs up to $20\%$ below random sampling when the property is targeted and more than $40\%$ below it when untargeted. In contrast, diversity policies show greater than random  RMSE-AUC improvement for all cases. NSGA-II Dis. + diversity most consistently selects informative materials for the most difficult future objectives. 

For the easier future objective, thermal conductivity, NSGA-II Dis. + diversity remains competitive. Across cases where these properties are targeted or untargeted, either individually or together with two additional properties, its mean performance remains above random sampling. These results strongly suggest that NSGA-II Dis. + diversity is robust for more difficult-to-predict properties while preserving competitive performance for easier objectives. In contrast, policies as HV $k$ medoids, which get on average comparative performance, can present extreme downgraded cases up to $20\%$ below random sampling.

\begin{figure*}[t]
    \centering
\includegraphics[width=1.\textwidth]{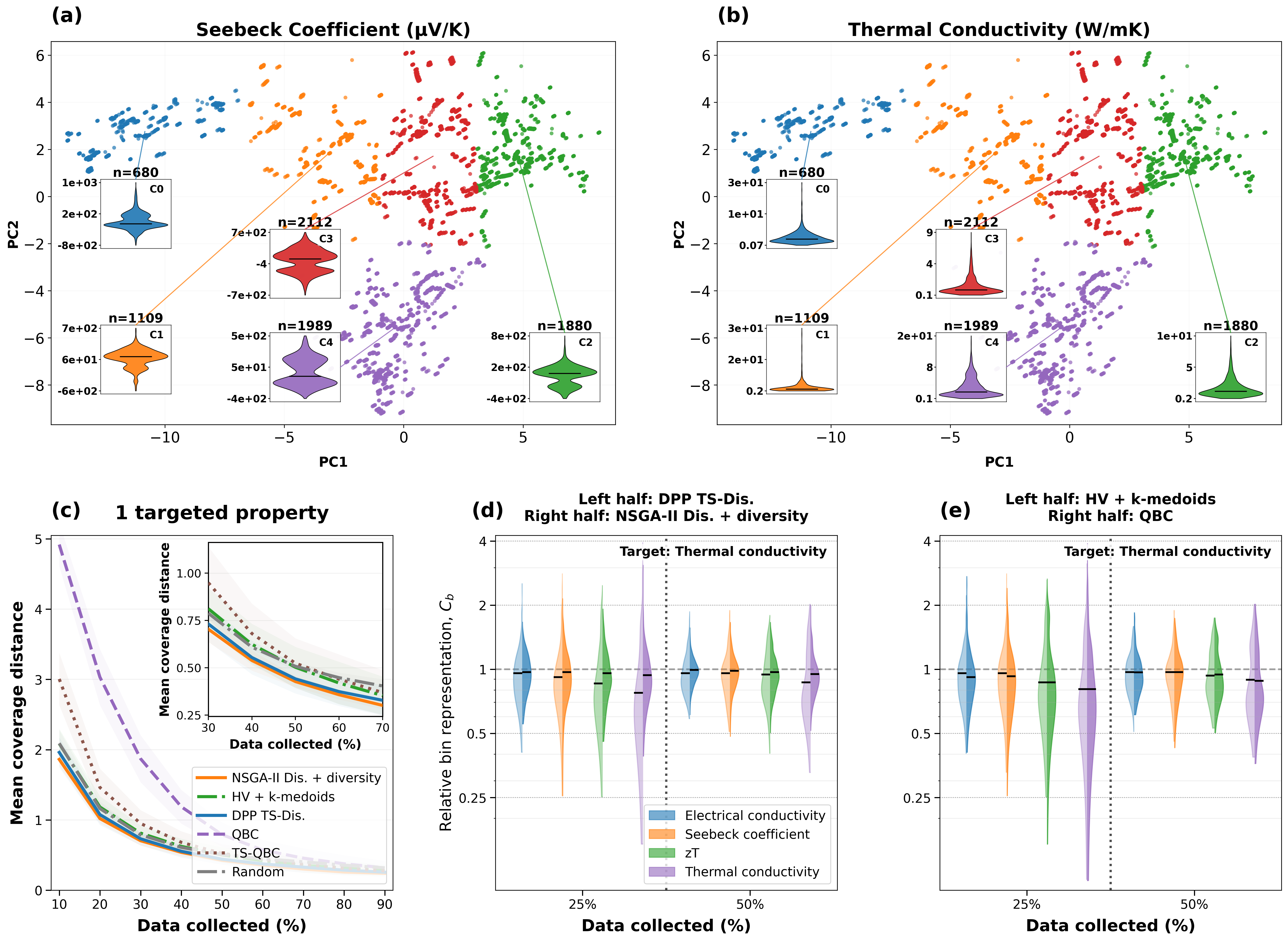}
\caption{ \textbf{(a,b)} First two principal components of sysTEm, with cluster-wise Seebeck coefficient and thermal conductivity distributions; numbers above violins give cluster sizes. \textbf{(c)} Mean feature-space distance to the remaining pool during one-property campaigns (lower is better). \textbf{(d,e)} Relative sampling distributions when targeting thermal conductivity (Eq.~\ref{relative_sampling}).}
\label{fig:violin_coverage_sysTEm}
\end{figure*}

The differences in prediction performance between easier and more difficult future objectives can be explained by the structure of the sysTEm dataset. The dataset contains dopant variants, so materials can be very close in feature space while having substantially different outcomes. This is visible in Fig.~\ref{fig:violin_coverage_sysTEm}(a,b), where materials are highly clustered and can exhibit broad property distributions. Under these conditions, disagreement alone becomes less discriminative. Closely related materials can obtain similar disagreement scores despite having different outcomes. QBC can therefore repeatedly select samples from the same local region. The corresponding distributions for electrical conductivity and $zT$ are shown in the Supplementary Information section \ref{improvement_dist}. Electrical conductivity shows a pattern similar to thermal conductivity, whereas Seebeck coefficient resembles $zT$.

The within-cluster property distributions further clarify the property-specific behavior. As shown in Fig.~\ref{fig:violin_coverage_sysTEm}(a), the Seebeck coefficient has broad distributions within clusters, making informative samples harder to identify using disagreement alone. In contrast, thermal conductivity often shows strongly skewed distributions with pronounced tails, as in clusters C1--C3 in Fig.~\ref{fig:violin_coverage_sysTEm}(b). In these cases, high-disagreement samples are more likely to be informative. This trend is consistent with the untargeted-property results. Thermal conductivity is among the easier properties to preserve, and electrical conductivity shows similar within-cluster distributions. Seebeck coefficient is more difficult to preserve when untargeted, while $zT$ shows a similar clustered distribution and performance pattern. These distributions can be found in the Supplmentary Information section \ref{Principal_components_of_datasets_and_properties_distribution}.

These results suggest that feature-space coverage also plays an important role. Fig.~\ref{fig:violin_coverage_sysTEm}(c) show the mean coverage distance between the unselected pool and the selected materials. For each unselected material, we calculate its distance to the nearest selected material and average these distances over the remaining pool. Lower values therefore indicate better feature-space coverage. Across one, two, and three targeted properties, QBC maintains the largest coverage distance throughout acquisition. Adding diversity substantially reduces this distance, explaining their good performance on targeted and untargeted properties. The differences among the diversity-aware policies are smaller, yet their ordering broadly follows their performance on untargeted properties.

With one targeted property, NSGA-II + diversity and DPP TS-Dis. provide the lowest coverage distances, with NSGA-II + diversity being slightly lower over most of the campaign Fig.~\ref{fig:violin_coverage_sysTEm}(c). HV+\(k\)-medoids follows closely. TS-Dis. has a higher distance and is often close to or above random sampling, while QBC has the poorest coverage. This ordering is mostly consistent with the untargeted results in Fig.~\ref{fig:Exp_ImprovementAll}(a).

Nevertheless, coverage by itself is not enough to understand difference between policies and robustness of NSGA-II Dis. + diversity. The complementary analysis shown in Fig.~\ref{fig:Exp_ImprovementAll}(d) show that NSGA-II policy tends to sample evenly from the outcome distributions while DPP TS-Dis. undersamples many ranges of property distribution. On the other hand, the sampling of HV +$k$ medoids property distribution is mostly the same as QBC. However, HV $k$ medoids shows more coverage than QBC. These results, show that diversifying in property disagreements mitigate the bias in  property distribution while diversifying in feature space ensures mitigating the concentration on feature space. Performing well on  both ensures strong performance on current objectives while mitigating future ones. And given that NSGA-II Dis. + diversity finds diverse Pareto frontier solution of disagreements and feature diversity, it achieves the best overall tradeoff of current and future objectives, even in the most challenging sampling scenarios. This general trend holds for all targeting cases and all other cases can be found in the Supplementary Information section \ref{Principal_components_of_datasets_and_properties_distribution}.

Overall, our results show that diversity-aware dataset construction is particularly important when sampling highly similar dopant materials with variable properties. Focused acquisition can degrade performance not only for future objectives but, in many cases, also for the properties currently being targeted. In contrast, diversity-aware sampling improves performance on targeted properties while reducing information loss for untargeted properties. Across the different diversity frameworks, NSGA-II Dis. + diversity provides the most robust overall strategy. It shows greater improvements over random sampling for all targeted properties while building datasets that remain informative for future objectives.
\section{Conclusions}
We developed and tested five diversity-aware dataset construction frameworks designed to improve performance for current objectives while reducing bias in future objectives. These approaches combine informative sampling with different forms of diversity and were compared against focused acquisition policies that consider only the current objectives.

Across all studied datasets, acquisition focused only on current targets can produce biased datasets that lose information about properties excluded from selection. This effect is especially strong in the experimental dopant dataset. There, focused policies can also underperform diversity-aware strategies on the targeted properties themselves.

Diversity mitigates this bias, but its effect depends on how it is introduced. Feature diversity alone does not consistently improve dataset quality across datasets. In contrast, promoting diversity in model disagreement can improve the balance between targeted and untargeted performance. Combining both forms of diversity provides the strongest overall balance. The benefit is greatest for similar dopant materials with highly variable outcomes. In this case, diversity-aware policies not only reduce degradation on future objectives but also improve performance on the properties used during acquisition. Thus, for datasets containing similar materials with different outcome behavior, diversity is important not only for preserving future usefulness but also for improving the dataset for current objectives.

Across all datasets, the NSGA-II Dis. + diversity framework introduced in this work provides the most consistent balance between current and future objectives. Even in the most challenging cases, it mitigates degradation on untargeted properties while maintaining targeted prediction performance above random sampling. This consistent behavior across both current and future objectives makes it the most robust strategy among those studied.

The NSGA-II Dis. + diversity framework supports dataset reuse when prediction objectives change. By preserving information relevant to untargeted properties, it can reduce the additional experiments or DFT calculations needed for future tasks. Such reuse potentially decreases cumulative computational costs and experimental resources across successive research objectives.

Overall, our results support a broader view of active learning for materials dataset construction. Acquisition should not only improve prediction for current objectives but also preserve information for properties that may become important in the future. Combining informative sampling with diversity provides a practical path to build datasets that remain useful for both current and future materials objectives.

\section{Methods}
\subsection{Datasets}\label{subsec1}
We selected five datasets that each contain at least three physically distinct and weakly linearly correlated properties. These datasets enable us to evaluate our framework in different data distribution settings. The experimental dataset contains clustered samples in feature space, as many compounds differ only by dopants. In contrast, the DFT datasets are broadly distributed. Furthermore, guaranteeing different uncorrelated outcomes ensures that performance on unconsidered properties in dataset construction cannot arise from linear inter-property correlations and instead reflects cross-objective generalization.

Four of the datasets are density functional theory (DFT) repositories: the 2018.06.01 (MP18) and 2021.11.10 (MP21) releases of the Materials Project, and the 2018.07.07 (JARVIS18) and 2022.12.12 (JARVIS22) releases of JARVIS. These DFT-based datasets provide three outcomes: electronic bandgap, elastic bulk modulus, and formation energy. Together, these properties probe different aspects of materials behavior — electronic structure, mechanical response, and thermodynamic stability. Furthermore, they are linearly weakly correlated (Pearson correlation values shown in the Appendix Fig.~\ref{fig:correlationDFT}), hence we avoid that sampling for one outcome ensures immediate informativeness for other targets.

Our analysis was restricted to materials for which bandgap, bulk modulus, and formation energy were jointly available, discarding all other entries. For JARVIS22 and MP21, we further refined the datasets by removing materials with formation energies exceeding 5 eV $\text{atom}^{-1}$, as such values correspond to highly unstable or nonphysical structures and are often associated with calculation artifacts rather than meaningful chemical behavior. Across all DFT datasets, we additionally verified that bulk modulus values were physically meaningful and retained only entries with bulk modulus greater than 0 GPa. Structural and compositional descriptors were then extracted using the Voronoi tessellation featurization~\cite{PhysRevB.96.024104} as implemented in \texttt{matminer}\cite{MATMINER}.\\

For the databases JARVIS18 and MP18, we did not directly use the originally reported structures or property values. Instead, we used the material identifiers from these older releases to query their corresponding entries in the most recent database versions, ensuring that all structures and labels reflected up-to-date calculations. This procedure mitigates inconsistencies introduced by database revisions.

The fifth dataset used in this work is the Systematically Verified Thermoelectric (sysTEm) dataset~\cite{Sparks_ThermoElectric}. It contains experimentally measured thermoelectric materials. The dataset was curated with doped and undoped materials. Evaluating our framework on this dataset allows us to assess whether acquisition strategies remain effective for experimental noise and heterogeneous measurements. Hence, our results demonstrate its robustness applicability for building informative datasets on experimental and computational campaigns.

Mixed or composite formulas were first parsed into weighted sub-compositions and combined into a single effective composition, ensuring correct stoichiometric treatment of multicomponent systems. Each resulting composition was then featurized using a comprehensive suite of composition-based descriptors from \texttt{matminer}, with measurement temperature included as an additional input feature. We removed redundant feature columns containing exclusively missing values, eliminated duplicated rows, and retained only four target outcome properties of: the thermoelectric figure of merit ($zT$), thermal conductivity, electrical conductivity, and the Seebeck coefficient. Additional properties, electrical thermal conductivity, lattice thermal conductivity and power factor that exhibit strong linear correlation with Thermal conductivity and $zT$ were excluded. This curation step ensures that performance on non-considered outcomes cannot be attributed to strong linear correlations among targets. The correlation matrices of the included features and the retained properties is provided in the Supplementary Information section \ref{Corr_matrices}.

For both the DFT datasets and the sysTEM experimental dataset, we removed highly linearly correlated features. We applied an iterative feature-removal procedure. Firstly, the absolute Pearson correlation matrix was  computed for all features and, for each feature, counted the number of other features with which it had $|\rho|>0.85$. The feature with the largest number of high-correlation relationships was then removed. The correlation matrix was recalculated after each removal, and the procedure was repeated until no pair of remaining features exceeded the correlation threshold. We applied this filtering to reduce redundancy in the feature space and to prevent any acquisition method from assigning disproportionate importance to directions represented by multiple highly correlated features.

The final number of curated entries used for the five data sets, number of features and potential targets for dataset construction is summarized in Table \ref{NumOfMaterials}. By evaluating performance jointly across DFT repositories and experimental data, we probe data sampling under two fundamentally different data regimes, with controlled and uncontrolled noise. Furthermore, this complementary design supports that our findings are independent of specific dataset structure or data distribution. 

\begin{table}[htbp]
\centering
\setlength{\tabcolsep}{4pt}
\begin{tabular}{c c c@{\hspace{0.03cm}}c}
\toprule
\textbf{Dataset} 
& \makecell{\textbf{Num.}\\\textbf{Materials}} 
& \makecell{\textbf{Num.}\\\textbf{Features}} 
& \makecell{\textbf{Potential}\\\textbf{Targets}} \\
\midrule

JARVIS18 & 18,964 & 141 
& \multirow{4}{*}{\makecell{
Bandgap,\\
Bulk\\modulus,\\
Formation\\energy\\
}} \\

JARVIS22 & 23,472 & 145 & \\
MP18     & 6,979  & 138 & \\
MP21     & 7,112  & 138 & \\ [0.8em]

\addlinespace
\midrule

\multirow{4}{*}[-1.5em]{sysTEm}
& \multirow{4}{*}[-1.5em]{7,771}
& \multirow{4}{*}[-1.5em]{161}
& \makecell{Electrical\\conductivity,} \\

& & & \makecell{Thermal\\conductivity,} \\

& & & \makecell{Seebeck\\coefficient,} \\

& & & \makecell{Figure of\\merit ($zT$)} \\

\bottomrule
\end{tabular}

\caption{Number of materials, features, and potential target properties used for each dataset after data curation.}
\label{NumOfMaterials}
\end{table}
\vspace{-1.25em}

\subsection{Active Learning Dataset Construction Policies}
\label{sec:acquisition_policies}
\subsubsection{Committee disagreement}
Let $\mathcal{U}_t$ be candidate pool at time iteration $t$, $\mathcal{O}$ denote the set of targeted properties and let $M=|\mathcal{O}|$. For candidate $\mathbf{s}\in\mathcal{U}_t$ and target $i\in\mathcal{O}$, committee disagreement was defined as

\begin{equation}
\Delta_i(\mathbf{s})
=
\left|
\hat{y}_{\mathrm{RF},i}(\mathbf{s})
-
\hat{y}_{\mathrm{XGB},i}(\mathbf{s})
\right|,
\label{eq:qbc_raw_disagreement}
\end{equation}
where $\hat{y}_{\mathrm{RF},i}$ and $\hat{y}_{\mathrm{XGB},i}$ are the predictions produced by the RF and XGB models trained for target $i$. This disagreement serves as a computational proxy for epistemic uncertainty: candidates for which the two model families disagree more strongly are treated as more informative.

Because the target properties can have different numerical scales, disagreement was normalized independently for each target over the current candidate pool:
\begin{equation}
q_i(\mathbf{s})
=
\frac{
\Delta_i(\mathbf{s})-
\min_{\mathbf{z}\in\mathcal{U}_t}\Delta_i(\mathbf{z})
}{
\max_{\mathbf{z}\in\mathcal{U}_t}\Delta_i(\mathbf{z})
-
\min_{\mathbf{z}\in\mathcal{U}_t}\Delta_i(\mathbf{z})
}.
\label{eq:normalized_disagreement}
\end{equation}
This normalization prevents targets with larger disagreement magnitudes from dominating multi-target acquisition scores. When the maximum and minimum disagreement were equal, the normalized disagreement for that target was set to zero.
\subsection{Acquisition policies}
We evaluated acquisition policies that differ in how they combine predictive disagreement, outcome-space trade-offs, stochastic exploration, and feature-space diversity.\\
\textbf{1. Query by Committee (QBC)}\\
The standard Query by Committee (QBC) policy aggregates normalized disagreement across the targeted properties:
\begin{equation}
a_{\mathrm{QBC}}(\mathbf{s})
=
\sum_{i\in\mathcal{O}} q_i(\mathbf{s}).
\label{eq:qbc_acquisition}
\end{equation}

At each iteration, candidates were ranked by $a_{\mathrm{QBC}}$, and the $|B|$ candidates with the largest scores were selected. This policy favors candidates with high total disagreement across the targeted properties but does not explicitly promote diversity within the selected batch.

\textbf{ 2. Non-dominated Sorting Genetic Algorithm II with disagreements and feature diversity (NSGA-II Dis. + diversity)}\\
For this policy, batch selection was formulated as a multi-objective optimization problem and solved using NSGA-II~\cite{996017}. Each NSGA-II individual represented a candidate batch $B\subseteq\mathcal{U}_t$ of the required acquisition size.
For target $i$, the mean normalized disagreement within batch $B$ was
\begin{equation}
\mathrm{AvgDis}_i(B)
=
\frac{1}{|B|}
\sum_{\mathbf{s}\in B} q_i(\mathbf{s}).
\label{eq:batch_average_disagreement}
\end{equation}
Feature diversity was evaluated both within the candidate batch and relative to the current training set. The within-batch diversity was defined as
\begin{equation}
\mathrm{div}_{\mathrm{within}}(B)=
1-
\frac{1}{|B|(|B|-1)}
\sum_{\substack{\mathbf{x}_i,\mathbf{x}_j\in B|i\neq j}}
\cos(\mathbf{x}_i,\mathbf{x}_j),
\label{eq:diversity_within}
\end{equation}
where $|B|(|B|-1)$ is the number of ordered pairs of distinct batch samples. A larger value indicates that the selected samples are more dissimilar to one another.
Diversity relative to the current training set was defined as
\begin{equation}
\mathrm{div}_{\mathrm{train}}(B)
=
\frac{1}{|B|}
\sum_{\mathbf{x}_i\in B}
\left[
1-
\frac{1}{|T_t|}
\sum_{\mathbf{x}_{j\in T_t}}
\cos(\mathbf{x}_i,\mathbf{x}_j)
\right]
\label{eq:diversity_train}
\end{equation}
This term favors candidates that are dissimilar, on average, to the samples already collected. The two components were combined as
\begin{equation}
\mathrm{Div}(B)
=
w_{\mathrm{train}}
\mathrm{div}_{\mathrm{train}}(B)
+
w_{\mathrm{within}}
\mathrm{div}_{\mathrm{within}}(B)
\label{eq:diversity_total}
\end{equation}
subject to
$w_{\mathrm{train}}+w_{\mathrm{within}}=1.$ NSGA-II optimized $M+1$ objectives: one $\mathrm{AvgDis}_i(B)$ objective for each targeted property and one $\mathrm{Div}(B)$ objective. The algorithm evolved candidate batches through selection, crossover, mutation, non-dominated sorting, and crowding-distance ranking.
After the final generation, one batch was selected from the resulting Pareto set $\mathcal{P}$. Each objective was first min--max normalized over $\mathcal{P}$. For diversity,
\[
\widetilde{\mathrm{Div}}(B)
=
\frac{
\mathrm{Div}(B)-
\min_{C\in\mathcal{P}}\mathrm{Div}(C)
}{
\max_{C\in\mathcal{P}}\mathrm{Div}(C)
-\min_{C\in\mathcal{P}}\mathrm{Div}(C)
},\]
and, for target $i$,
\begin{dmath}
\widetilde{\mathrm{AvgDis}}_i(B)
=
\frac{
\mathrm{AvgDis}_i(B)
-\min_{C\in\mathcal{P}}\mathrm{AvgDis}_i(C)
}{
\max_{C\in\mathcal{P}}\mathrm{AvgDis}_i(C)
-
\min_{C\in\mathcal{P}}\mathrm{AvgDis}_i(C)
}.
\label{eq:normalized_average_disagreement}
\end{dmath}
The final batch was chosen using equal-weight scalarization:
\begin{equation}
B_t^{*}
=
\arg\max_{B\in\mathcal{P}}
\frac{
\widetilde{\mathrm{Div}}(B)
+
\sum_{i\in\mathcal{O}}
\widetilde{\mathrm{AvgDis}}_i(B)
}{
M+1
}.
\label{eq:nsgaii_with_diversity}
\end{equation}
Thus, the selected batch lies on the Pareto front and provides a balanced compromise between target-specific disagreement and feature-space diversity.

\textbf{3. Non-dominated Sorting Genetic
Algorithm-II with disagreements (NSGA-II Dis.)}\\
To isolate the effect of explicit feature-space diversity, we also evaluated an NSGA-II policy based only on target-specific disagreement. Candidate batches were optimized using the $M$ objectives $\mathrm{AvgDis}_i(B)$, without including $\mathrm{Div}(B)$.

After the final generation, the selected batch was
\begin{equation}
B_t^{*}
=
\arg\max_{B\in\mathcal{P}}
\frac{1}{M}
\sum_{i\in\mathcal{O}}
\widetilde{\mathrm{AvgDis}}_i(B),
\label{eq:nsgaii_without_diversity}
\end{equation}
where $\widetilde{\mathrm{AvgDis}}_i(B)$ is defined in Eq.~\ref{eq:normalized_average_disagreement}.
Although this policy does not enforce feature-space diversity, NSGA-II uses crowding distance to maintain a distributed set of solutions in the multi-objective disagreement space. It therefore promotes trade-offs among target-specific disagreement objectives but does not directly encourage chemical or feature-space coverage.

\textbf{4. Hypervolume of disagreements (HV)}\\
For multi-target campaigns, normalized disagreement was also combined using a hypervolume-based acquisition score. For candidate $\mathbf{s}$, the score was defined as the volume of the axis-aligned region dominated by its normalized disagreement vector relative to reference point $\mathbf{r}=0$. Therefore, 
\begin{equation}
a_{\mathrm{HV}}(\mathbf{s})
=
\prod_{i=1}^{M}q_i(\mathbf{s}).
\end{equation}
At each iteration, all candidates were ranked by $a_{\mathrm{HV}}$, and the $B$ candidates with the largest scores were selected. Because the score is multiplicative, it favors candidates for which the committee disagrees simultaneously across all targeted properties.

\textbf{5. Hypervolume of disagreements with $k$-medoids (HV + $k$-medoids)}
This policy first used $a_{\mathrm{HV}}$ to identify an informative subset of the candidate pool. The number of retained candidates was
\begin{equation}
K_t
=
\min\left(
|\mathcal{U}_t|,
\left\lceil
\rho|\mathcal{U}_t|
\right \rceil \right)
\quad
\rho=0.10
\label{eq:hv_top_fraction}
\end{equation}
The retained subset was
\begin{equation}
\mathcal{V}_{\text{HV},t}
=\operatorname{Top}_{K_t}
\left\{
a_{\mathrm{HV}}(\mathbf{s}) \big|
 \mathbf{s}\in\mathcal{U}_t
\right\}.
\label{eq:hv_retained_subset}
\end{equation}
The feature vectors in $\mathcal{V}_{\text{HV},t}$ were standardized using statistics computed from the combined collected and candidate data and were subsequently normalized to unit Euclidean norm. Partitioning around medoids was then applied to $\mathcal{V}_{\text{HV},t}$ using $k=B$. The selected batch was
\begin{equation}
B_t^{*}
=
\underset{\mathcal{M}\subseteq \mathcal{V}_{\text{HV},t},
|\mathcal{M}|=B}
{\text{arg min}}=
\sum_{\mathbf{x}\in \mathcal{V}_{\text{HV},t}}
\min_{\mathbf{m}\in\mathcal{M}}
\left|
\widetilde{\mathbf{x}}-
\widetilde{\mathbf{m}}
\right|_2,
\label{eq:hv_kmedoids}
\end{equation}
where $\widetilde{\mathbf{x}}$ denotes a preprocessed feature vector. The hypervolume stage identifies candidates with high simultaneous disagreement, whereas $k$-medoids selects a representative and feature-diverse batch from that informative subset.

\textbf{6. Thompson Sampling using disagreements (TS-Dis.)}\\
We introduced stochastic exploration by perturbing the normalized QBC disagreement scores. This procedure is inspired by Thompson sampling but does not construct a predictive posterior over the target functions.

For candidate $\mathbf{s}$, normalized disagreement was averaged across the $M$ targeted properties:

\begin{equation}
u(\mathbf{s})
=
\frac{1}{M}
\sum_{i=1}^{M}q_i(\mathbf{s}).
\label{eq:mean_qbc_score}
\end{equation}
For the $N_t=|\mathcal{U}_t|$ uncollected candidates, these scores were collected in
\begin{equation}
\mathbf{u}_t
=
\left[
u(\mathbf{s}_1),
\ldots,
u(\mathbf{s}_{N_t})
\right]^{\mathsf T}.
\end{equation}
We then defined the Gaussian approximation
\begin{equation}
\widetilde{\mathbf{u}}_t
\sim
\mathcal{N}
\left(
\mathbf{u}_t,
\boldsymbol{\Sigma}_t
\right),
\end{equation}
with diagonal covariance
\begin{equation}
\boldsymbol{\Sigma}_t
=
\tau^2
\operatorname{diag}
\left[
u(\mathbf{s}_1)^2,
\ldots,
u(\mathbf{s}_{N_t})^2
\right].
\label{eq:diagonal_ts_covariance}
\end{equation}
The diagonal approximation assumes zero correlation between candidate acquisition values and avoids constructing a dense candidate--candidate covariance matrix. Consequently, sampling from the joint approximation is equivalent to independently drawing
\begin{equation}
\widetilde{u}(\mathbf{s})
\sim
\mathcal{N}
\left(
u(\mathbf{s}),
\tau^2u(\mathbf{s})^2
\right)
\label{eq:thompson_perturbation}
\end{equation}
for every candidate. The perturbation parameter was set to $\tau=1$. Candidates with larger disagreement therefore had both larger expected acquisition values and greater sampling variance.

Because disagreement is non-negative, negative draws were clipped to zero:
\begin{equation}
a_{\mathrm{TS}}(\mathbf{s})
=
\max\left(
\widetilde{u}(\mathbf{s}),0
\right).
\label{eq:thompson_acquisition}
\end{equation}
A new acquisition vector was sampled at every active-learning iteration. Candidates were ranked by $a_{\mathrm{TS}}$, and the $B$ candidates with the largest sampled values were selected.

\textbf{7. Determinantal point process Thompson Sampling using disagreements (DPP TS-Dis.)}\\
This policy combined the stochastic acquisition values in Eq.~\ref{eq:thompson_acquisition} with feature-space diversity through a determinantal point process (DPP). The Thompson-inspired scores served as candidate quality values.

First, the top-scoring candidate subset was retained. Its size was determined as in equation \ref{eq:hv_top_fraction}  and the retained subset is
\begin{equation}
\mathcal{V}_{\text{TS},t}
=
\operatorname{Top}_{K_t}
\left\{
a_{\mathrm{TS}}(\mathbf{s}):
\mathbf{s}\in\mathcal{U}_t
\right\}
\end{equation}

The feature vectors within $\mathcal{V}_{\mathrm{TS},t}$ were standardized and then min--max normalized. Pairwise feature similarity was computed using a radial basis function kernel:
\begin{equation}
S_{ij}
=
\exp\left(
-\frac{
\left\|
\widetilde{\mathbf{x}}_i-
\widetilde{\mathbf{x}}_j
\right\|_2^2
}{
2\sigma^2
}
\right),
\label{eq:dpp_rbf_similarity}
\end{equation}
where $\sigma$ was set to the square root of the median nonzero squared pairwise distance among retained candidates, with a fallback value of $1$ when all distances were zero.
The retained Thompson-inspired scores were min--max normalized and shifted by $\varepsilon=10^{-10}$ to obtain positive quality weights $q_i$. The DPP kernel was defined as
\begin{equation}
L_{ij}
=
q_i S_{ij} q_j.
\label{eq:dpp_kernel}
\end{equation}
For numerical stability, the kernel was symmetrized and a diagonal term $\varepsilon I$ was added, yielding $\widehat{L}$. The batch was selected using a greedy approximation to the fixed-cardinality DPP maximum a posteriori problem:
\begin{equation}
B_t^{*}
\approx
\underset{
B\subseteq\mathcal{V}_{\text{TS},t},\,
|B|=b
}{
\text{arg max}
}
\det\left(\widehat{L}_B\right),
\label{eq:dpp_map}
\end{equation}
where $b$ is the prescribed batch size and $\widehat{L}_B$ is the kernel submatrix indexed by $B$. For the unregularized kernel,
\begin{equation}
\det(L_B)
=
\left(\prod_{i\in B}q_i^2\right)\det(S_B),
\end{equation}
which expresses the contributions of candidate quality and feature-space diversity. Starting from an empty batch, each greedy step added the remaining candidate that maximized the increase in the log-determinant, until $b$ candidates had been selected. If no candidate yielded a finite log-determinant with positive determinant sign, the remaining candidate with the largest kernel diagonal was selected. The quality weights favored candidates with large Thompson-inspired disagreement scores, whereas the determinant penalized redundancy and promoted diversity within the batch.

\subsection{Performance measure}
\label{subsec_performance_measure}
At each active-learning iteration $t$, we trained a Random Forest regressor and an XGBoost regressor for each outcome using the current training set $T_t$. Predictive performance was evaluated using the root mean squared error RMSE until the candidate pool was fully sampled. To compare each acquisition policy with random sampling over the full campaign, we calculated the area under the RMSE learning curve for targeted and untargeted properties, $\mathrm{AUC}_{\mathrm{RMSE}}
=
\int \mathrm{RMSE}(p)dp$, where $p$ is the fraction of the data collected. The improvement respect to random sampling of policy $\pi$ for an
individual property case $c$ was then calculated as

\begin{equation}
\mathrm{I}_{\pi,c}
=
\left(
1-
\frac{
{\mathrm{AUC}}_{\mathrm{RMSE},\pi,c}
}{
{\mathrm{AUC}}_{\mathrm{RMSE},\mathrm{Random}}
}
\right)\times100\%,
\end{equation}

where ${\mathrm{AUC}}_{\mathrm{RMSE},\pi,c}$ and
${\mathrm{AUC}}_{\mathrm{RMSE},\mathrm{Random}}$ denote the
RMSE-AUC for policy $\pi$ and random sampling,
respectively. Each evaluated property was classified as targeted if it was included in the
acquisition objective and untargeted otherwise. For a campaign targeting
$m$ properties, each target combination therefore
contained $m$ targeted properties and $4-m$ untargeted properties for the experimental datasets construction and $3-m$ for DFT datasets. 
\section*{Author contributions}
R.E.C. and J.H.S. conceived and designed the project. J.H.S. supervised the project. R.E.C and A.S.D. conducted the experiments. R.E.C. drafted the manuscript. Hongchen W., K.L. and A.B.D. provided technical support for the diversity policy construction. R.Z, Y.K. and Hao W. helped to conduct experiments in the Digital Research Alliance of Canada supercomputer.  All authors discussed the results and reviewed the manuscript.
\section*{Conflicts of interest}
The authors declare no conflicts of interest.

\section*{Data and code availability}

The data and code supporting this work will be made publicly available upon acceptance. 

\section*{Acknowledgments}
The authors acknowledge financial support from
the Natural Sciences and Engineering Research
Council of Canada (NSERC) Alliance grants
(ALLRP 601812-24), the National Research
Council of Canada’s Critical Battery Materials
Initiative (CBMI-002-1). The research was also,
in part, made possible thanks to funding provided
to the University of Toronto’s Acceleration Consortium by the Canada First Research Excellence
Fund (CFREF-2022-00042). Ashley Dale is a postdoctoral fellow at the University of Toronto in the Eric and Wendy Schmidt AI in Science Postdoctoral Fellowship Program, a program of Schmidt Sciences.

\begingroup
\footnotesize
\bibliography{main} 
\bibliographystyle{rsc} 
\endgroup

\clearpage

\onecolumn
\appendix
\clearpage
\section{Correlation Matrices}\label{Corr_matrices}

\begin{figure*}[!ht]
        \centering
\includegraphics[width=1\linewidth]{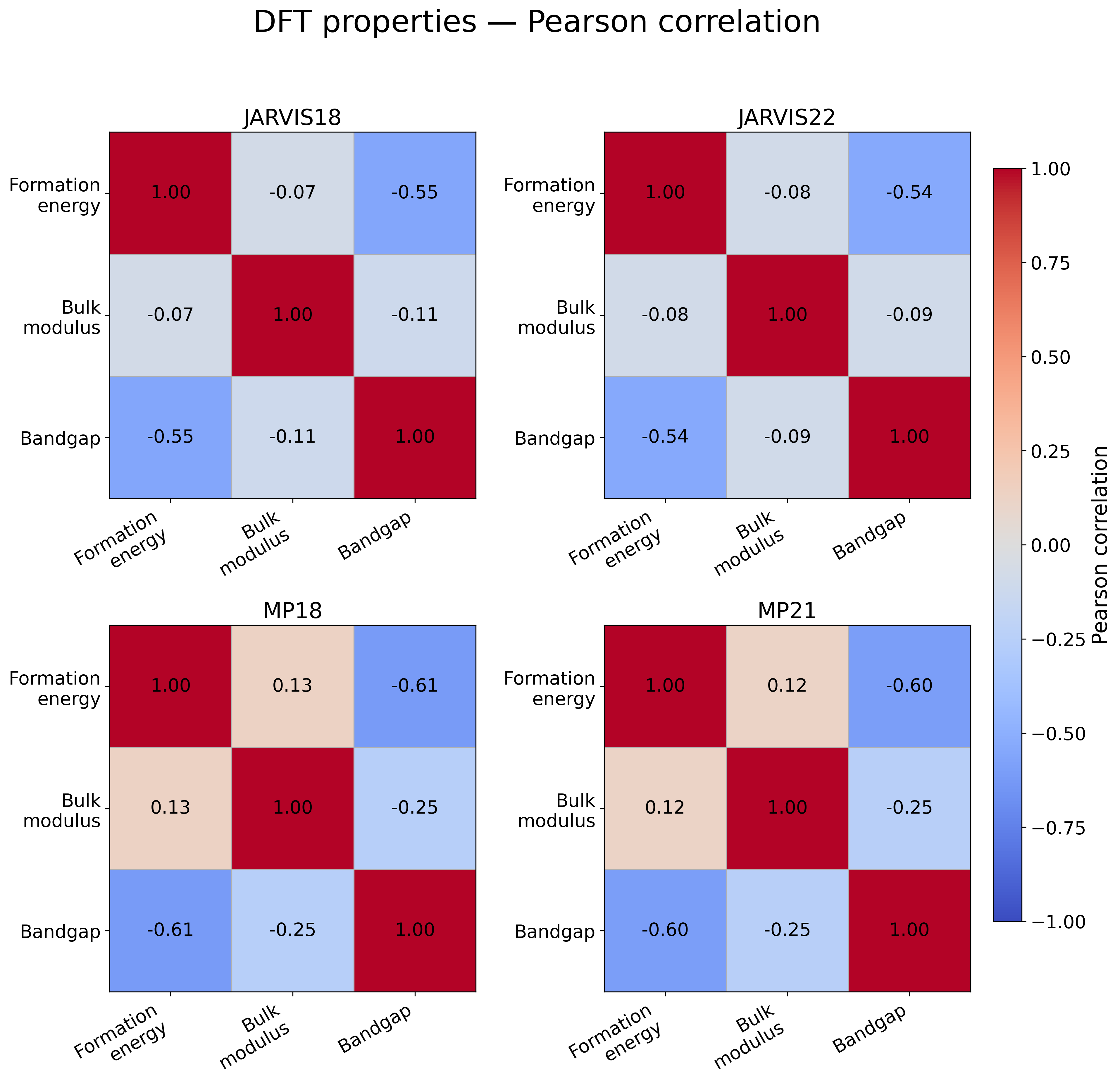}
\caption{
   \textbf{Pearson correlations among the properties used for dataset construction in the DFT datasets.} The magnitude of the correlation between any pair of distinct properties does not exceed 0.61.  
    }
\label{fig:correlationDFT}
\end{figure*}

\begin{figure*}[!ht]
        \centering
\includegraphics[width=1.\linewidth]{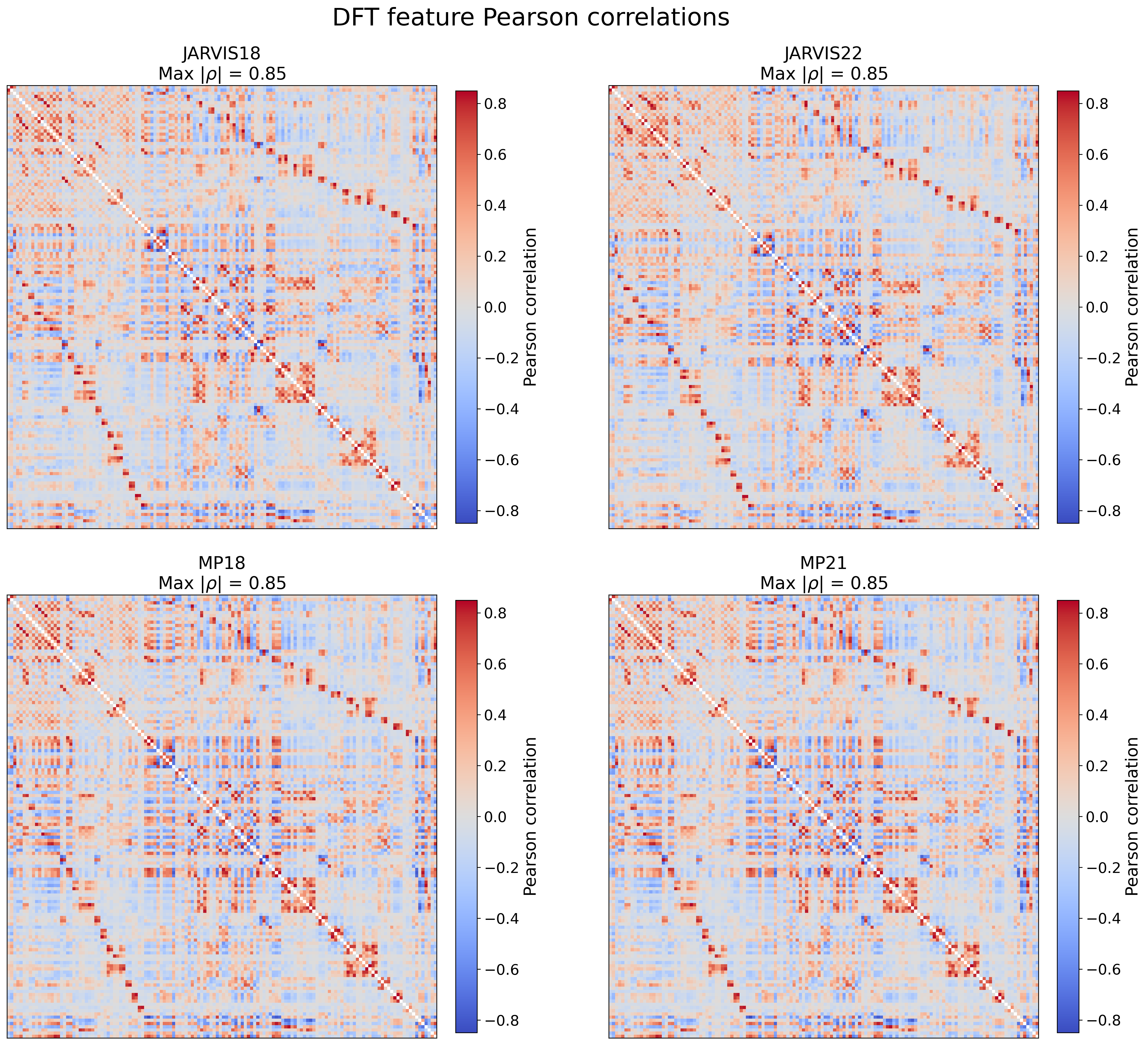
}
\caption{
\textbf{Pearson correlations among the features in the curated DFT datasets.} During data curation, correlated features were removed so that the magnitude of the Pearson correlation between any pair of distinct features did not exceed 0.85. Diagonal elements are excluded to highlight the maximum correlation between distinct features.
}
\label{fig:feature_DFTCorrelations}
\end{figure*}

\begin{figure*}[!ht]
        \centering
\includegraphics[width=0.6\linewidth]{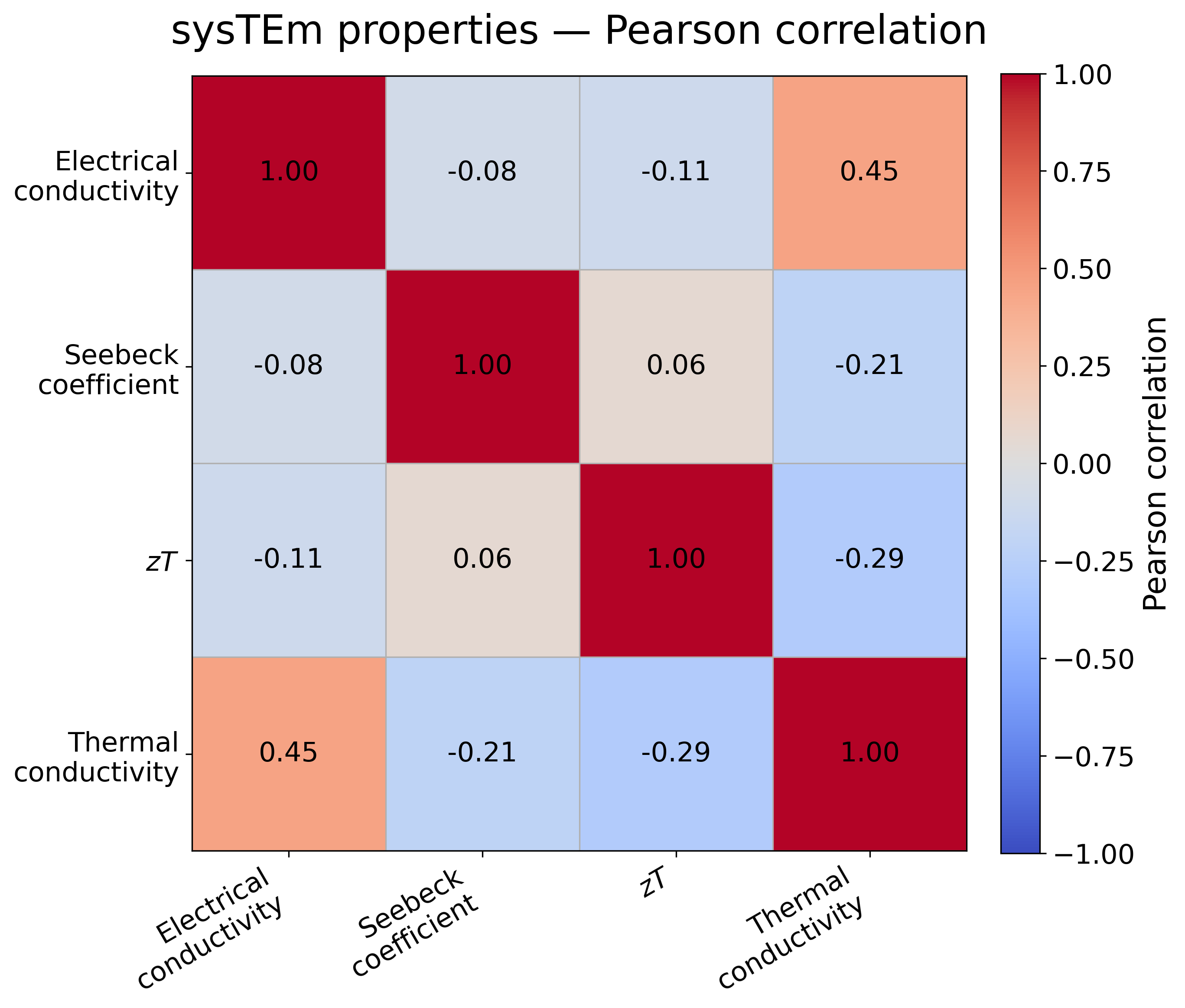}
\caption{
   \textbf{Pearson correlations among the properties of interest for dataset construction in the sysTEm dataset.} The magnitude of the correlation between any pair of distinct properties does not exceed 0.45.  
    }
\label{fig:correlationOutcomePropsysTEm}
\end{figure*}

\begin{figure*}[!ht]
        \centering
\includegraphics[width=0.6\linewidth]{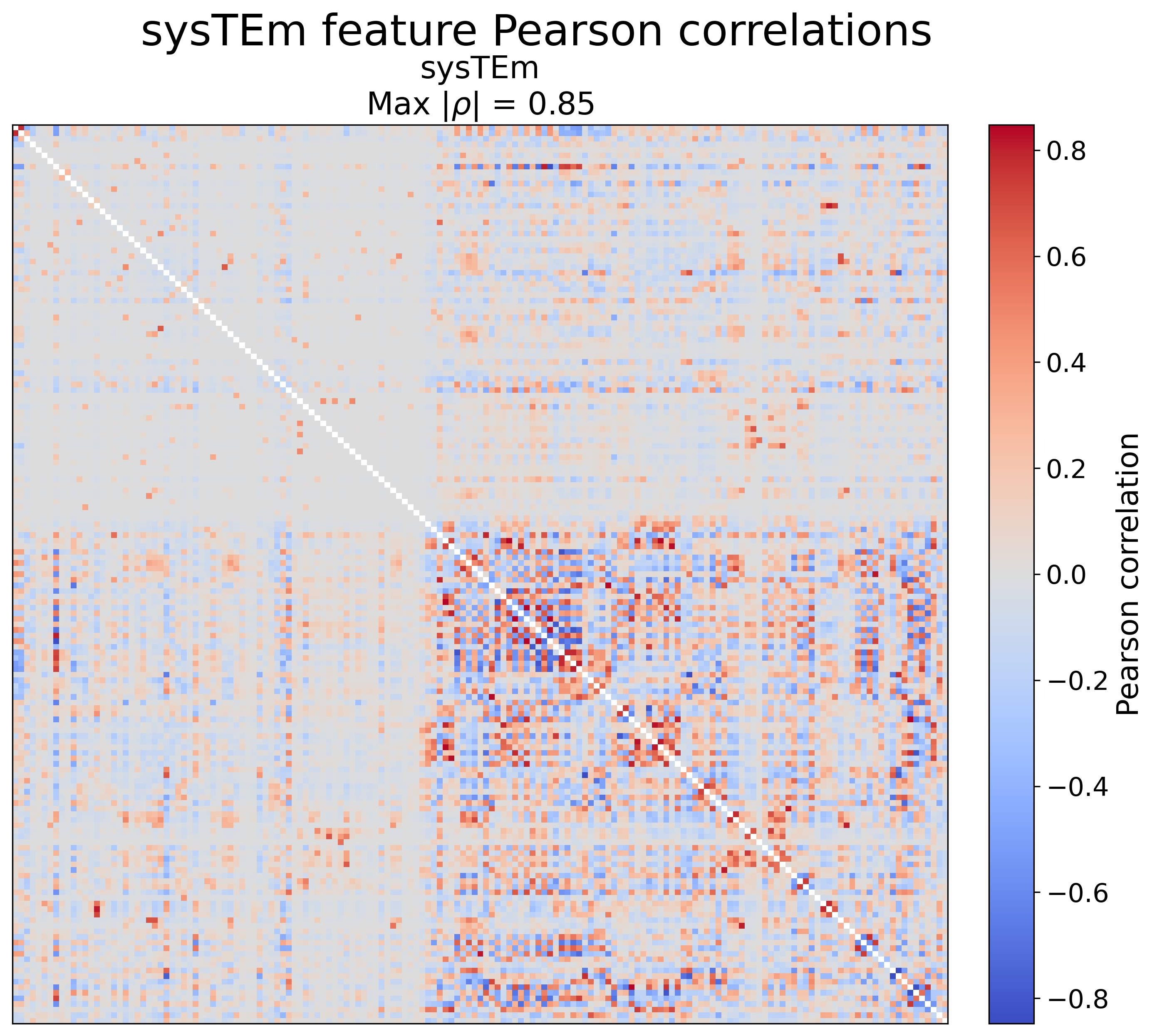}
\caption{
\textbf{Pearson correlations among the features in the curated sysTEm dataset.} During data curation, correlated features were removed so that the magnitude of the Pearson correlation between any pair of distinct features did not exceed 0.85. Diagonal elements are excluded to highlight the maximum correlation between distinct features.
}
\label{fig:correlationFeaturesysTEm}
\end{figure*}

\FloatBarrier
\section{Improvement distributions}
\label{improvement_dist}

\begin{figure*}[!ht]
        \centering
\includegraphics[width=0.9\linewidth]{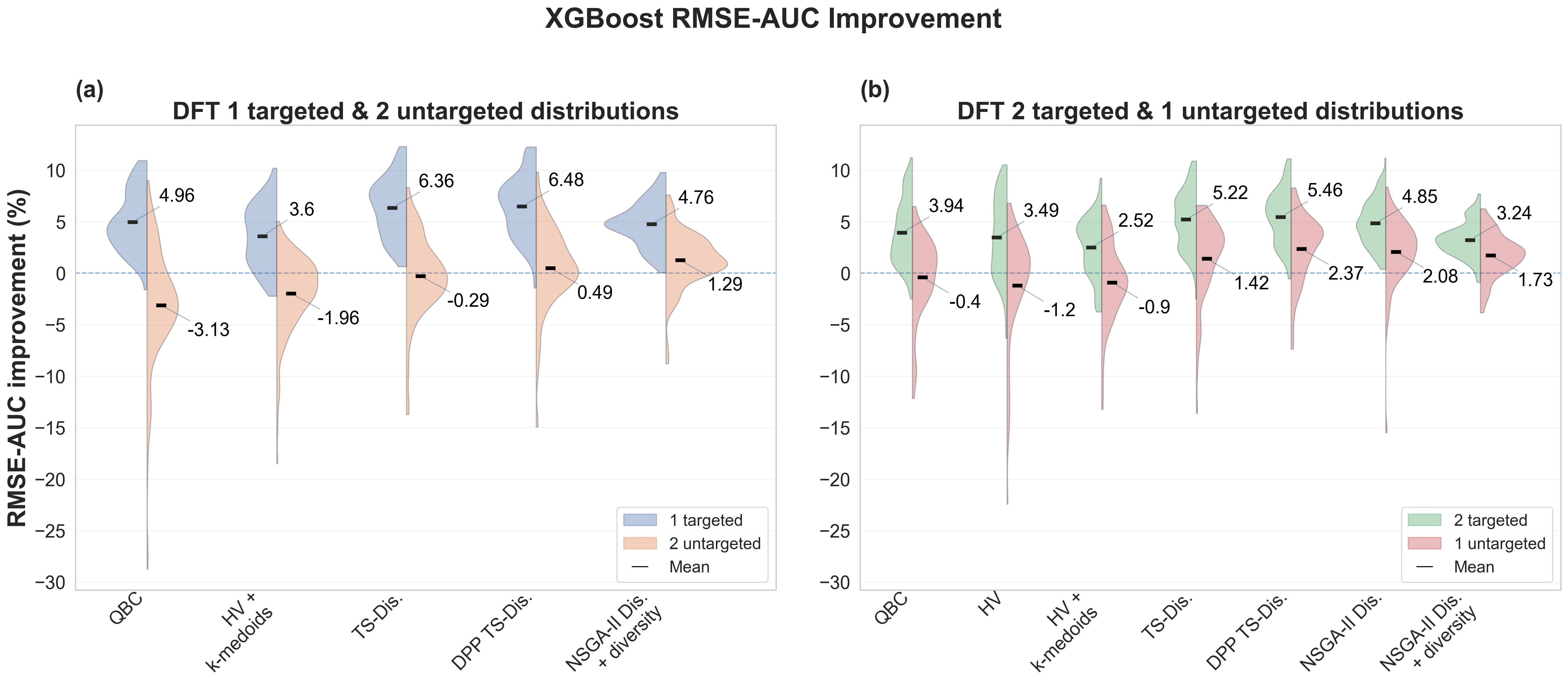}
\caption{
\textbf{Distributions of RMSE-AUC improvement for one- and two-property acquisition cases across all DFT datasets using XGBoost.}
    }
\label{fig:DFT_RF_XGBoost_2_2corr}
\end{figure*}

\begin{figure*}[!ht]
        \centering
\includegraphics[width=1.\linewidth]{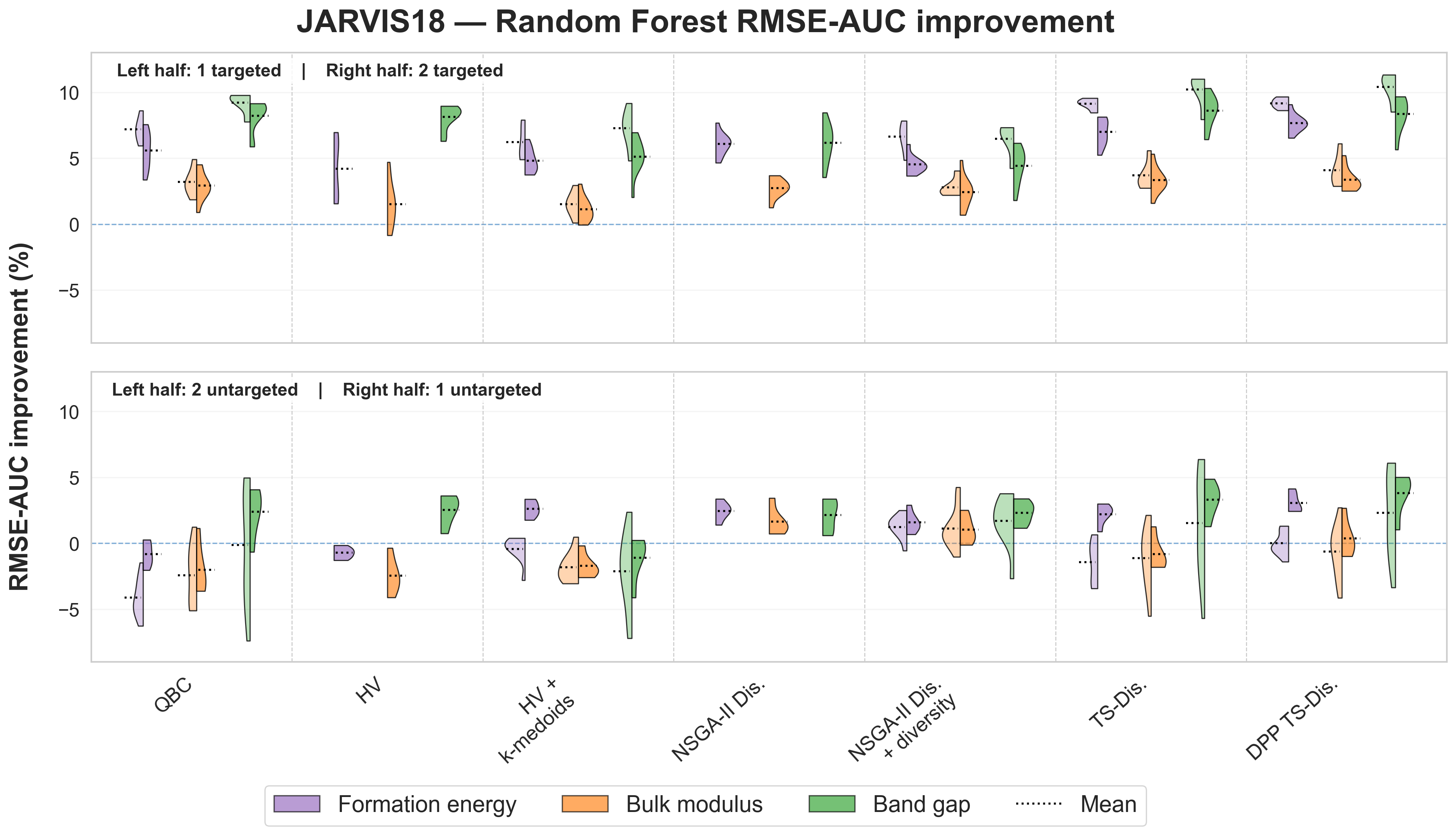}
\caption{\textbf{Distributions of RMSE-AUC improvement for random forest model predicting formation energy, bulk modulus, and band gap in the JARVIS18 dataset.}}
\end{figure*}

\begin{figure*}[!ht]
        \centering
\includegraphics[width=1.\linewidth]{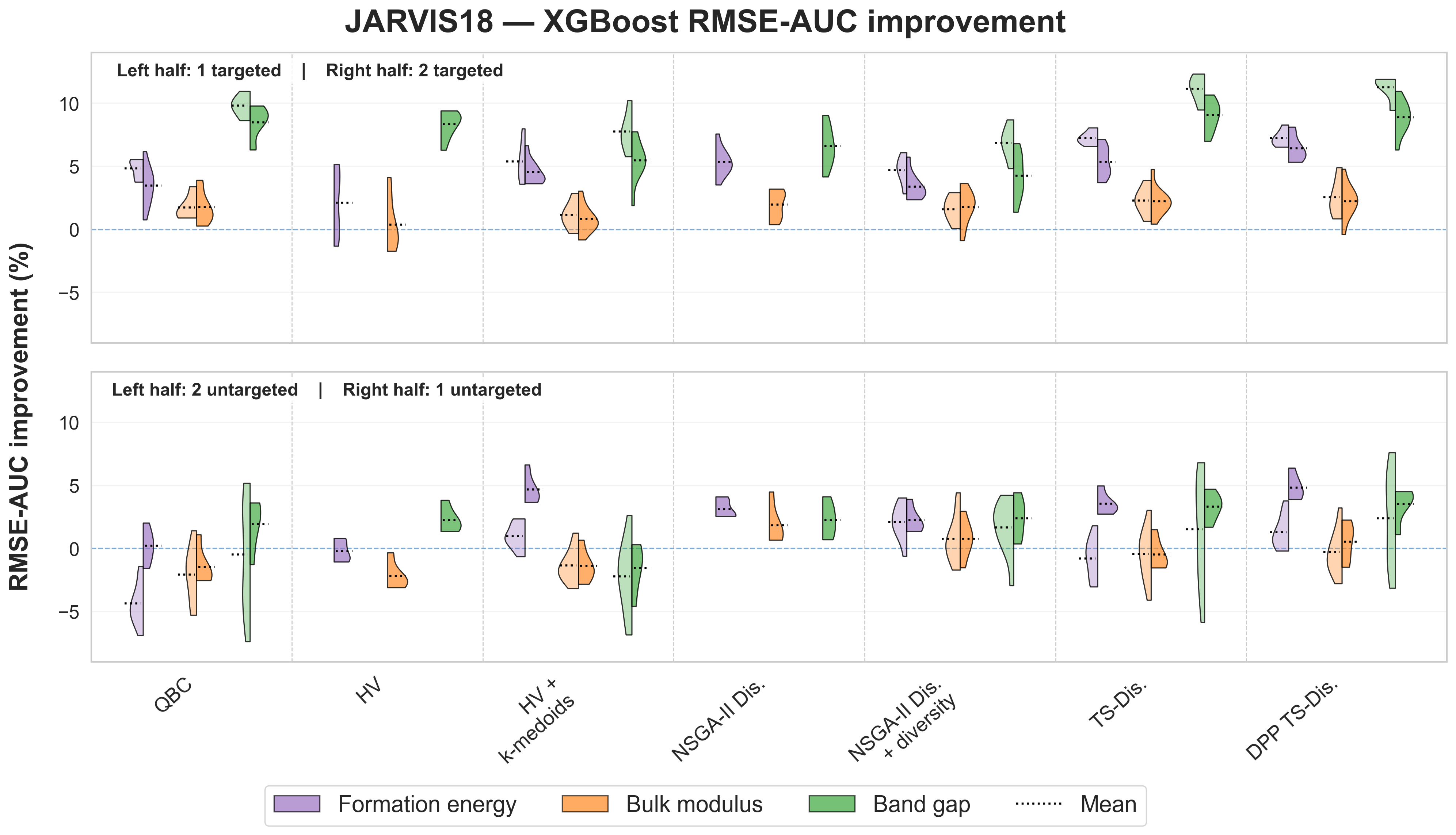}
\caption{\textbf{Distributions of RMSE-AUC improvement for XGBoost model predicting formation energy, bulk modulus, and band gap in the JARVIS18 dataset.}}
\end{figure*}
\begin{figure*}[!ht]
        \centering
\includegraphics[width=.9\linewidth]{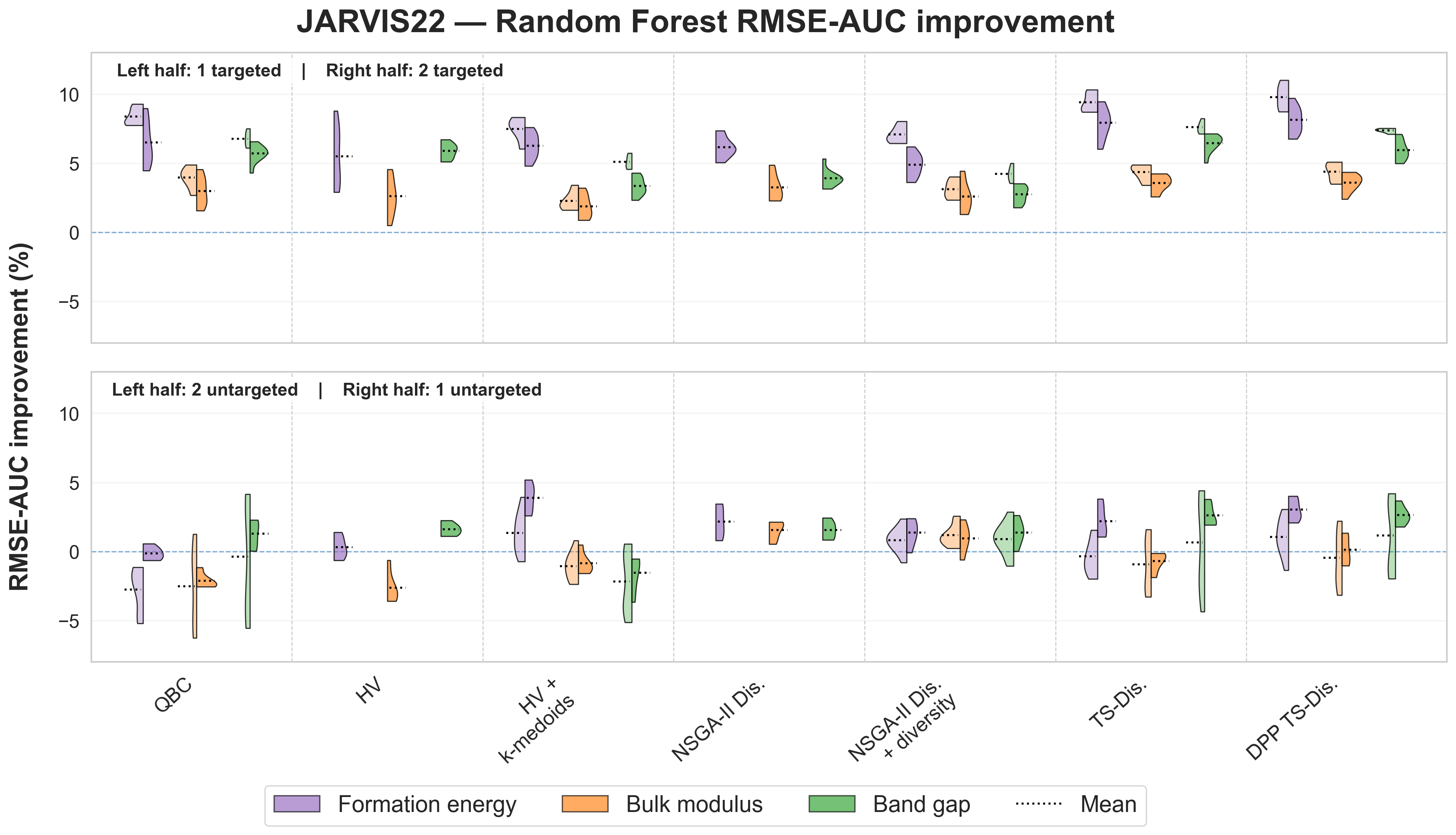}
\caption{\textbf{Distributions of RMSE-AUC improvement for random forest model predicting formation energy, bulk modulus, and band gap in the JARVIS22 dataset.}}
\end{figure*}

\begin{figure*}[!ht]
        \centering
\includegraphics[width=.9\linewidth]{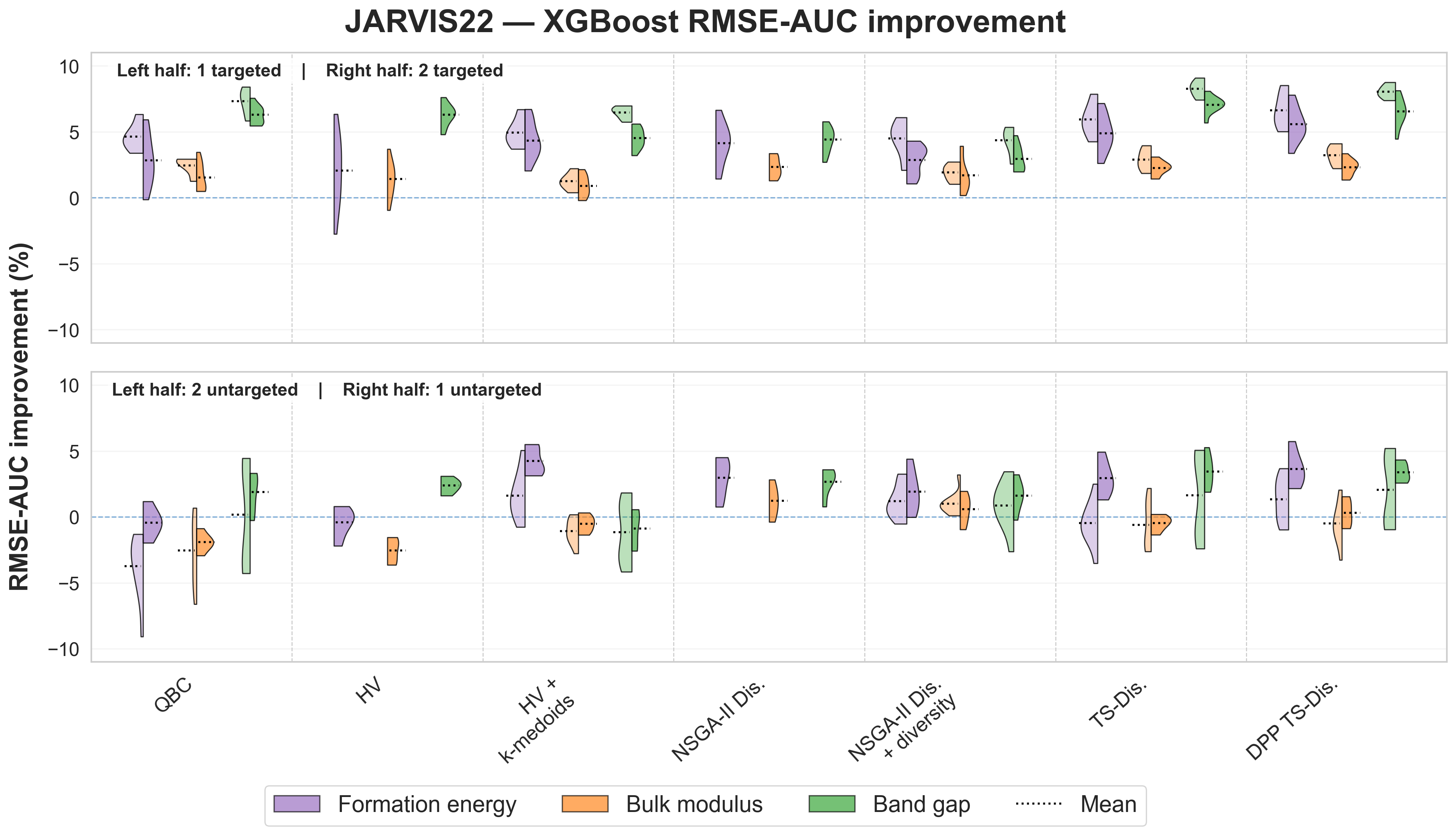}
\caption{\textbf{Distributions of RMSE-AUC improvement for XGBoost model predicting formation energy, bulk modulus, and band gap in the JARVIS22 dataset.}}
\end{figure*}

\begin{figure*}[!ht]
        \centering
\includegraphics[width=1.\linewidth]{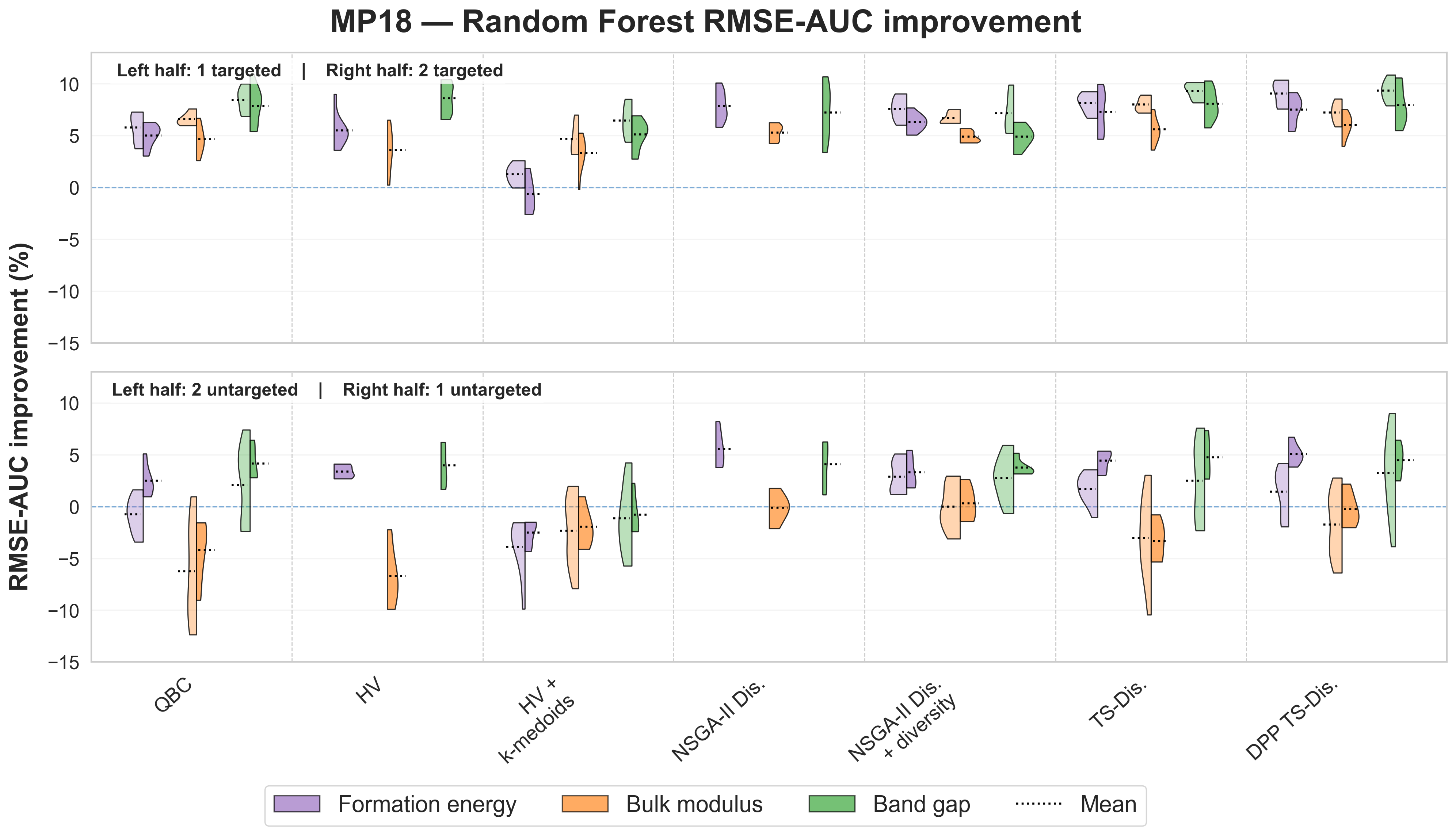}
\caption{\textbf{Distributions of RMSE-AUC improvement for random forest model predicting formation energy, bulk modulus, and band gap in the MP18 dataset.}}
\end{figure*}

\begin{figure*}[!ht]
        \centering
\includegraphics[width=.9\linewidth]{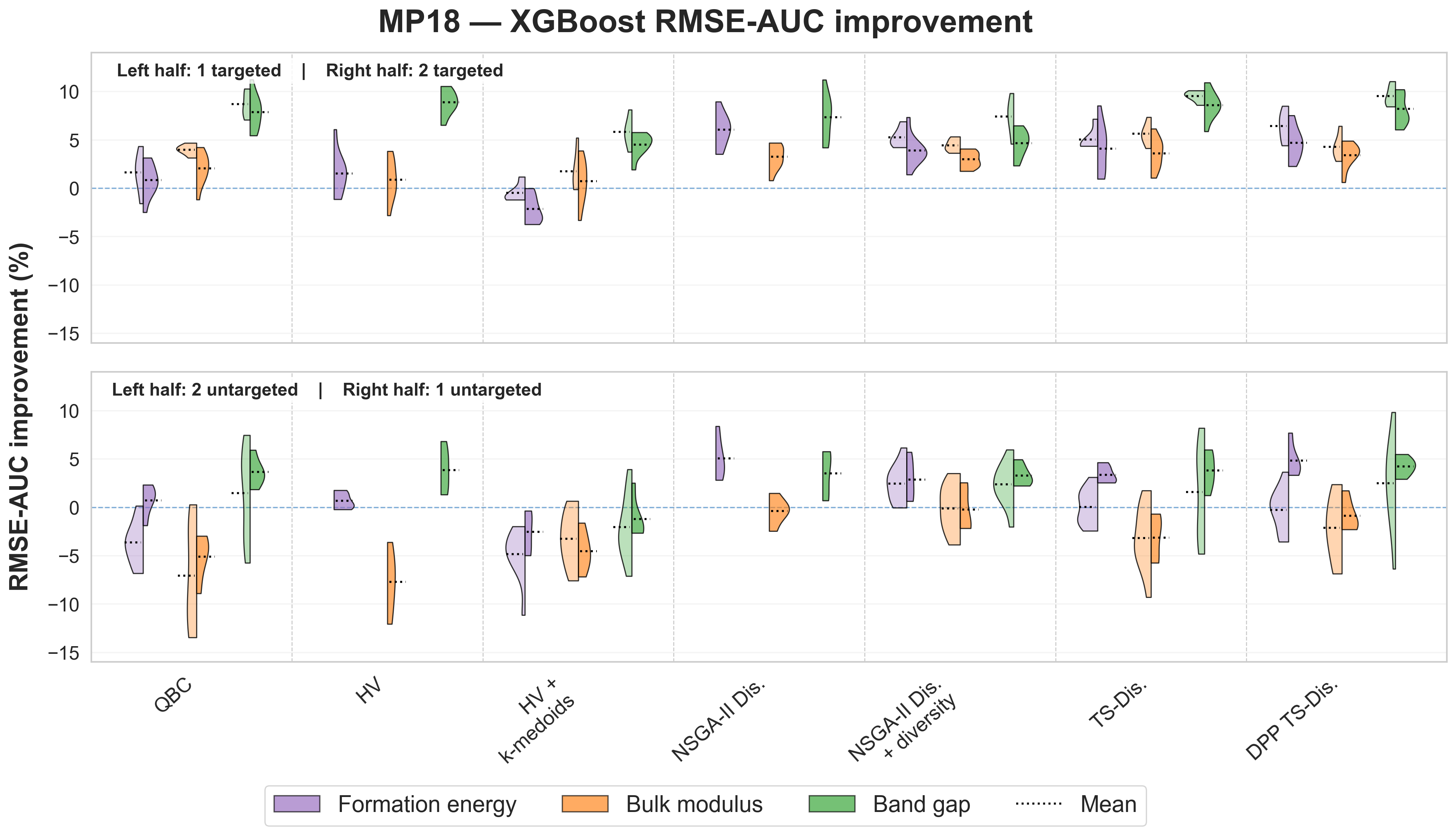}
\caption{\textbf{Distributions of RMSE-AUC improvement for XGBoost model predicting formation energy, bulk modulus, and band gap in the MP18 dataset.}}
\end{figure*}

\begin{figure*}[!ht]
        \centering
\includegraphics[width=.9\linewidth]{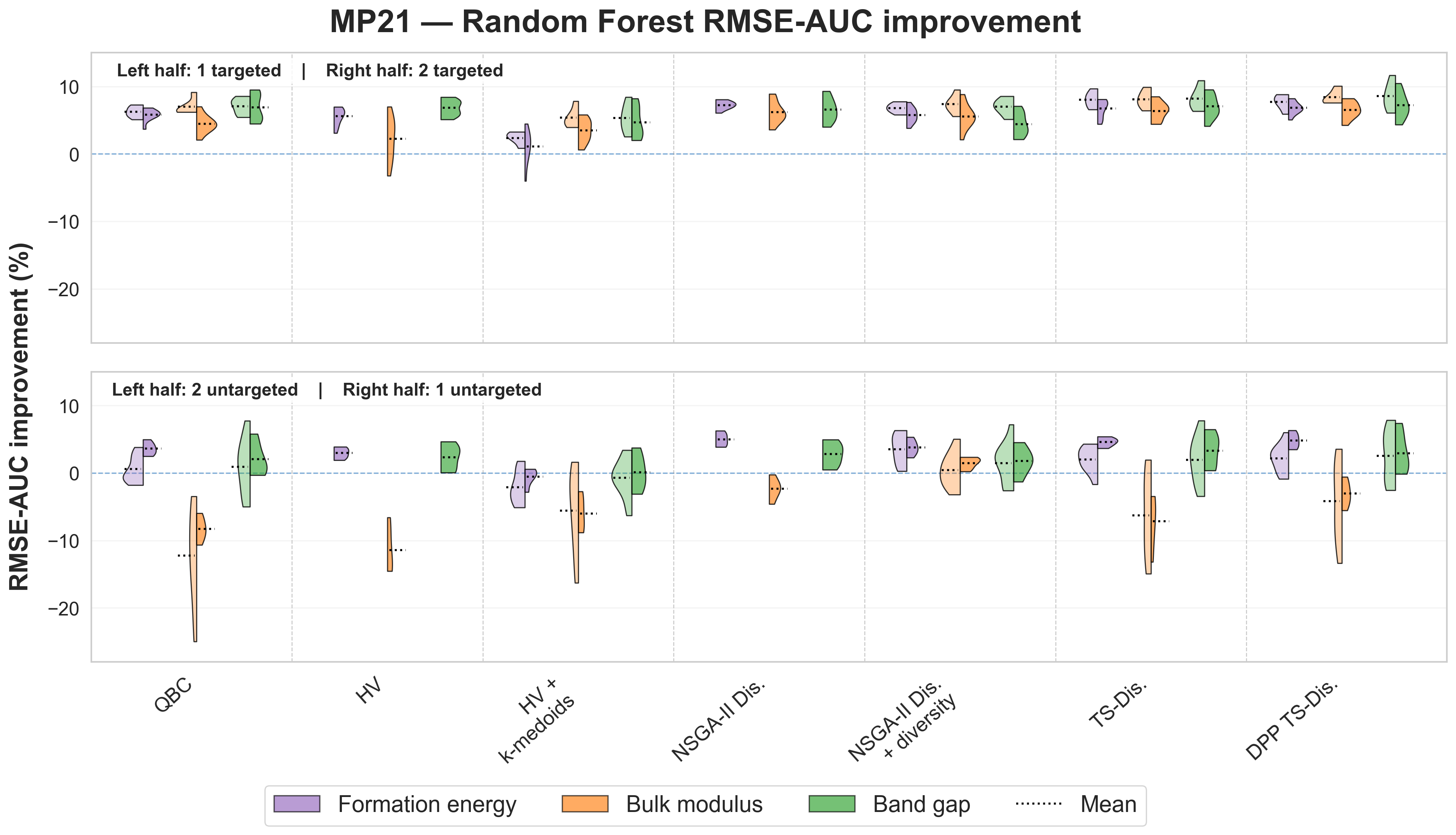}
\caption{\textbf{Distributions of RMSE-AUC improvement for random forest model predicting formation energy, bulk modulus, and band gap in the MP21 dataset.}}
\end{figure*}

\begin{figure*}[!ht]
        \centering
\includegraphics[width=.9\linewidth]{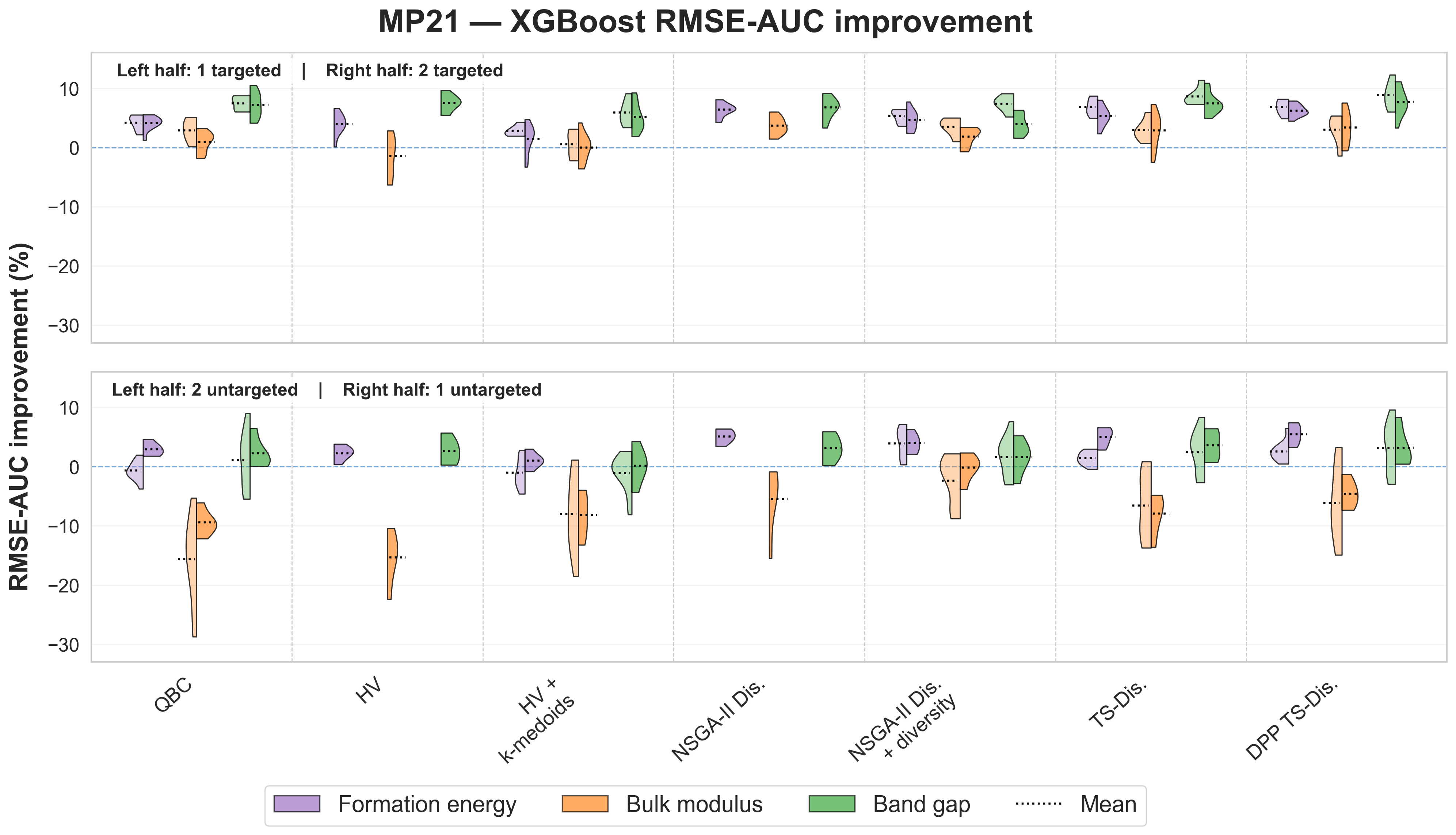}
\caption{\textbf{Distributions of RMSE-AUC improvement for XGBoost model predicting formation energy, bulk modulus, and band gap in the MP21 dataset.}}
\end{figure*}

\begin{figure*}[!ht]
        \centering
\includegraphics[width=1\linewidth]{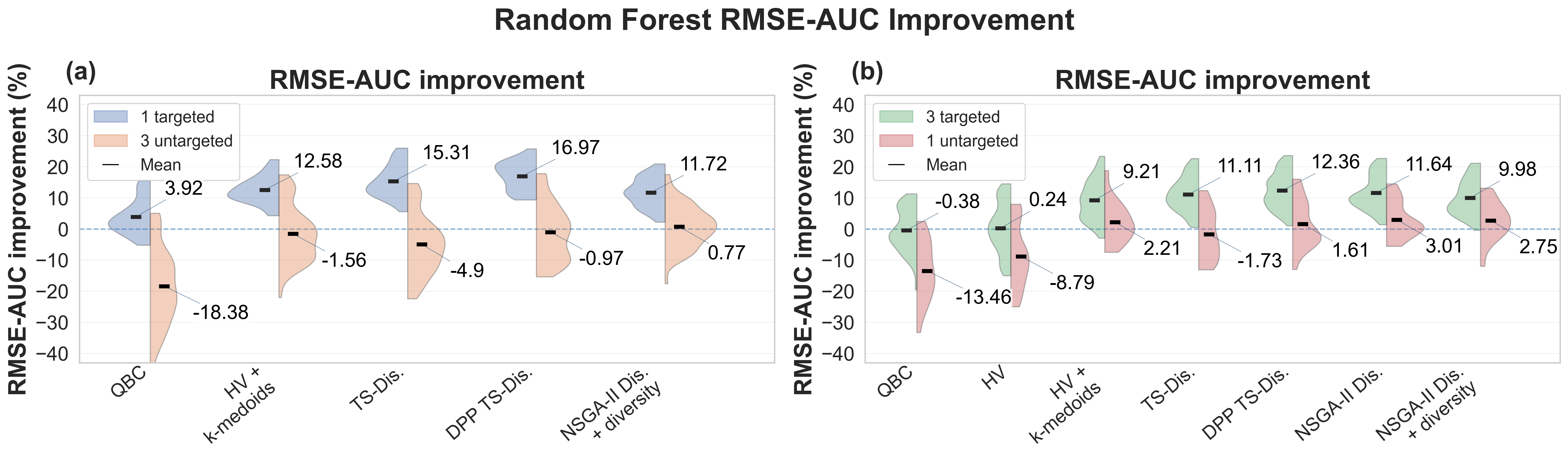}
\caption{
\textbf{Distributions of RMSE-AUC improvement for one- and three-property acquisition cases with sysTEm dataset using Random Forest.}
    }
\label{fig:sysTEm_1_3_XGBoosts}
\end{figure*}

\begin{figure*}[!ht]
        \centering
\includegraphics[width=1.\linewidth]{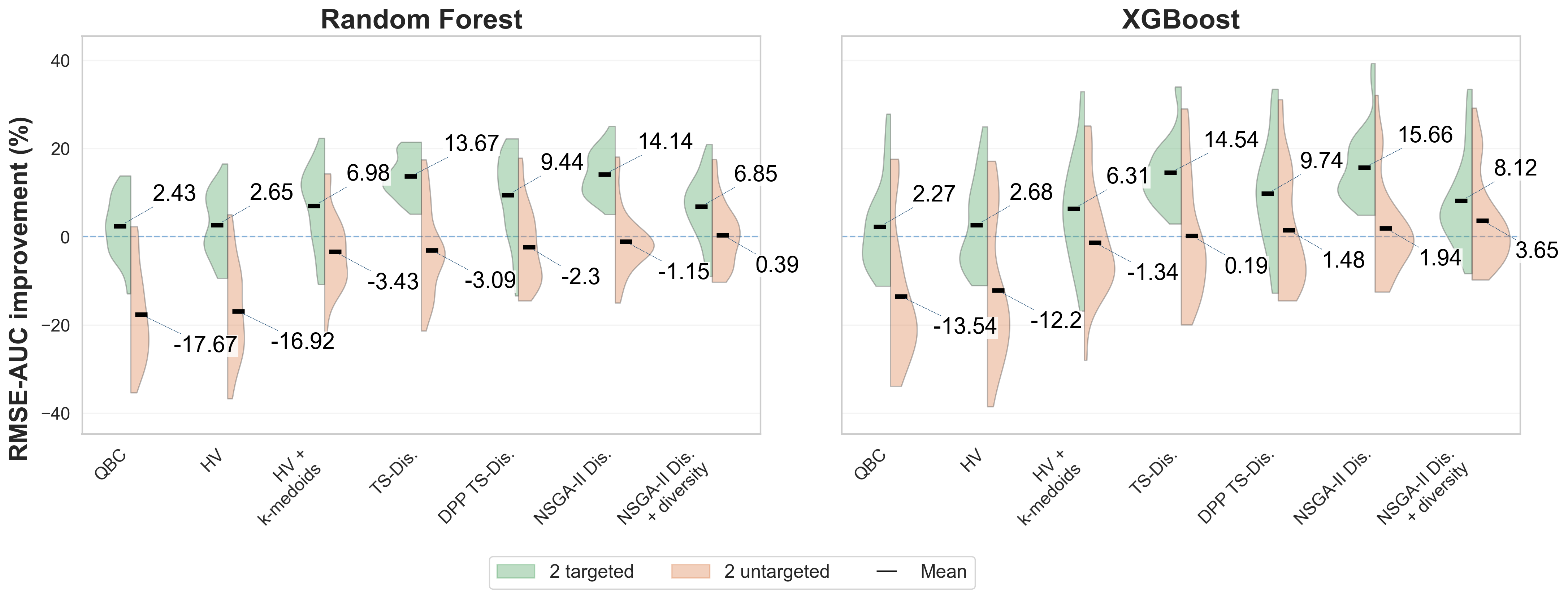}
\caption{
\textbf{Distributions of RMSE-AUC improvement for two property acquisition cases with sysTEm dataset using Random Forest and XGBoost.}
    }
\label{fig:sysTEm_2targ_2untarg}
\end{figure*}

\begin{figure*}[!ht]
        \centering
\includegraphics[width=.85\linewidth]{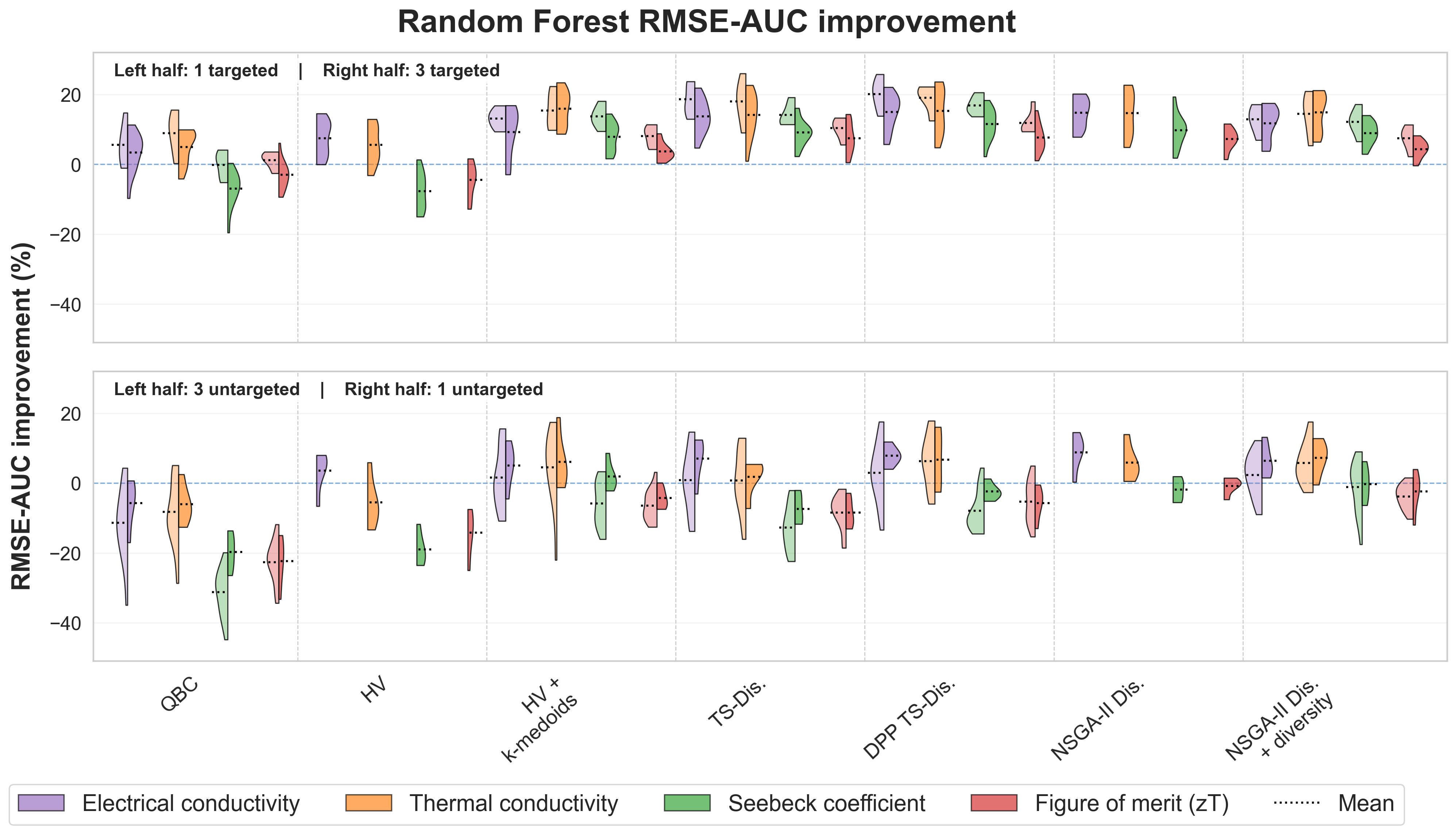}
\caption{
Distributions of targeted and untargeted RMSE-AUC improvement when targeting one or three properties, using random forest model to predict electrical conductivity, thermal conductivity, the Seebeck coefficient, and the thermoelectric figure of merit (zT) in the sysTEm dataset.
    }
\label{fig:sysTEm_3_1_RandomForest}
\end{figure*}

\begin{figure*}[!ht]
        \centering
\includegraphics[width=0.85\linewidth]{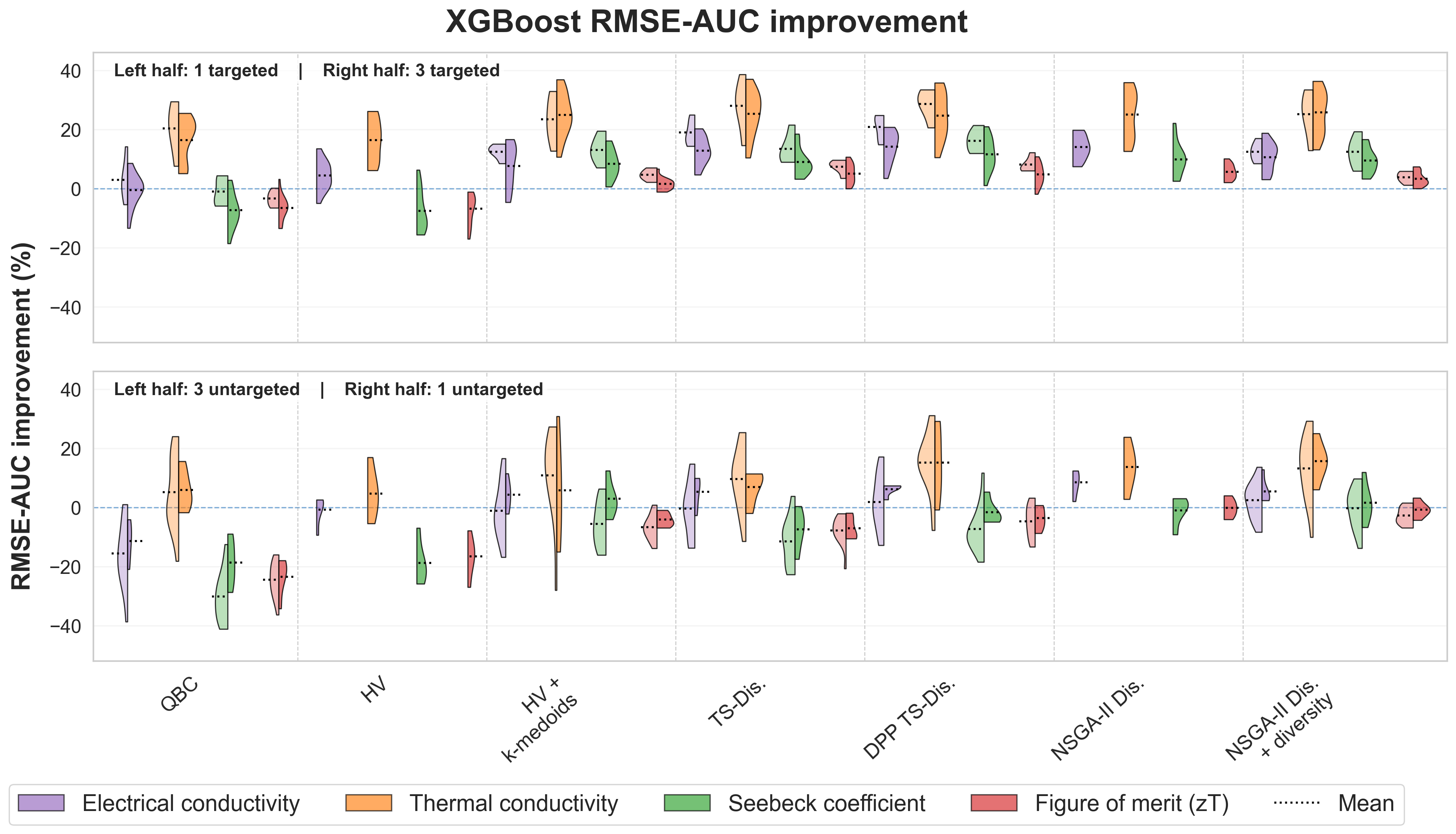}
\caption{
Distributions of targeted and untargeted RMSE-AUC improvement when targeting one or three properties, using XGBoost model to predict electrical conductivity, thermal conductivity, the Seebeck coefficient, and the thermoelectric figure of merit (zT) in the sysTEm dataset.
    }
\label{fig:sysTEm_3_1_XGBoosts}
\end{figure*}

\begin{figure*}[!ht]
        \centering
\includegraphics[width=0.9\linewidth]{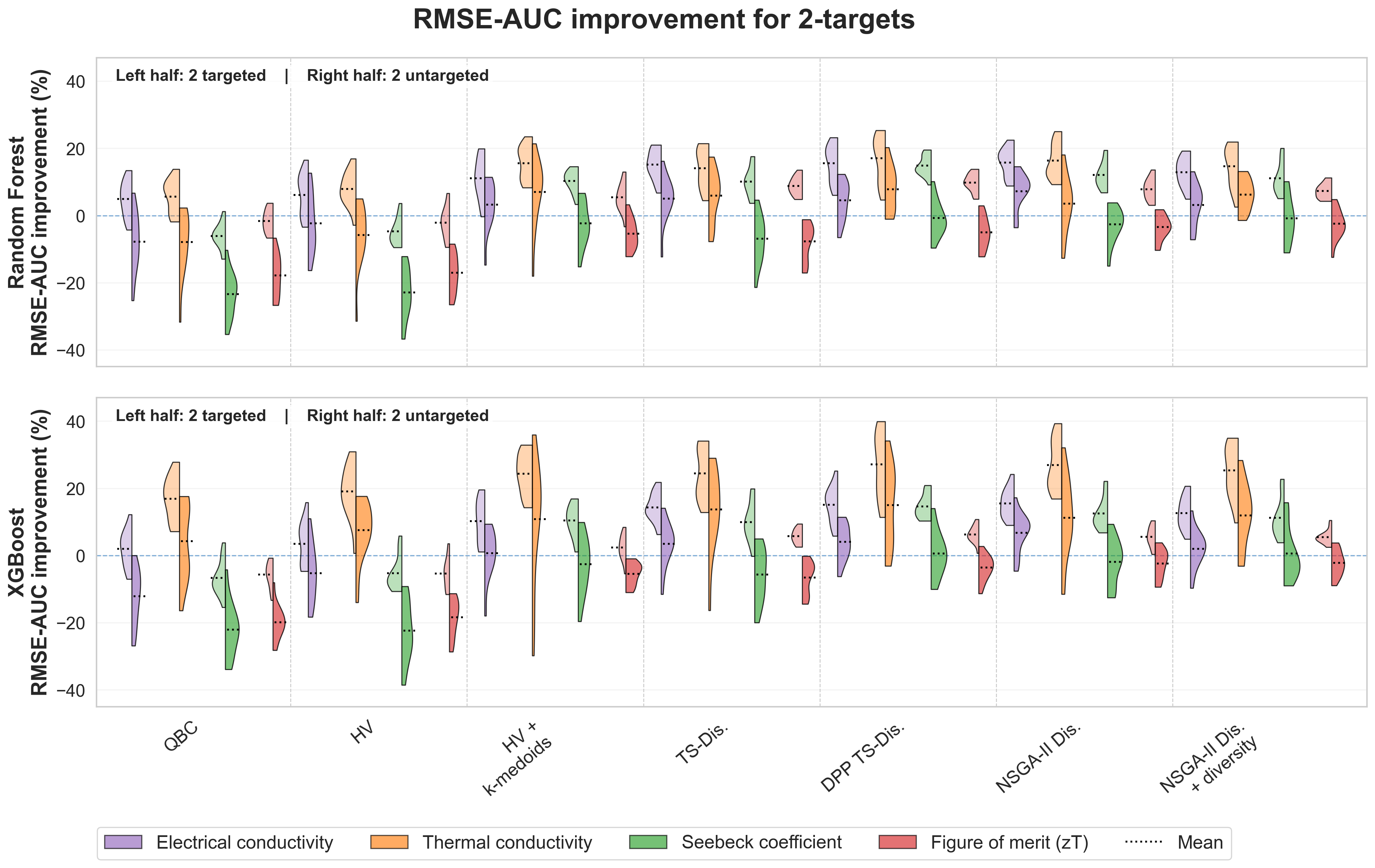}
\caption{Distributions of targeted and untargeted RMSE-AUC improvement when targeting two properties, using random forest and XGBoost models to predict electrical conductivity, thermal conductivity, the Seebeck coefficient, and the thermoelectric figure of merit (zT) in the sysTEm dataset. 
    }
\label{fig:sysTEm_2Targets}
\end{figure*}

\FloatBarrier
\section{Principal components of datasets and properties distribution}
\label{Principal_components_of_datasets_and_properties_distribution}

\begin{figure*}[!ht]
        \centering
\includegraphics[width=\linewidth]{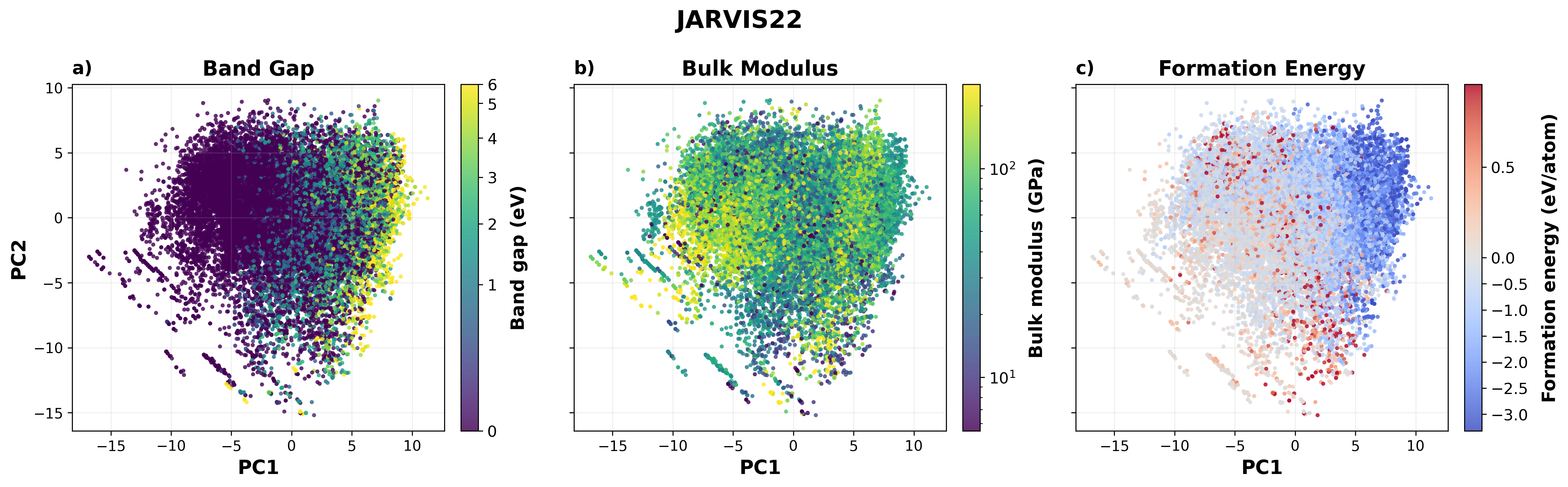}
\caption{
 Properties distribution of JARVIS22 shown in the first two principal components of feature space.
    }
\label{fig:DistributionOutcomesJARVIS22}
\end{figure*}

\begin{figure*}[!ht]
        \centering
\includegraphics[width=\linewidth]{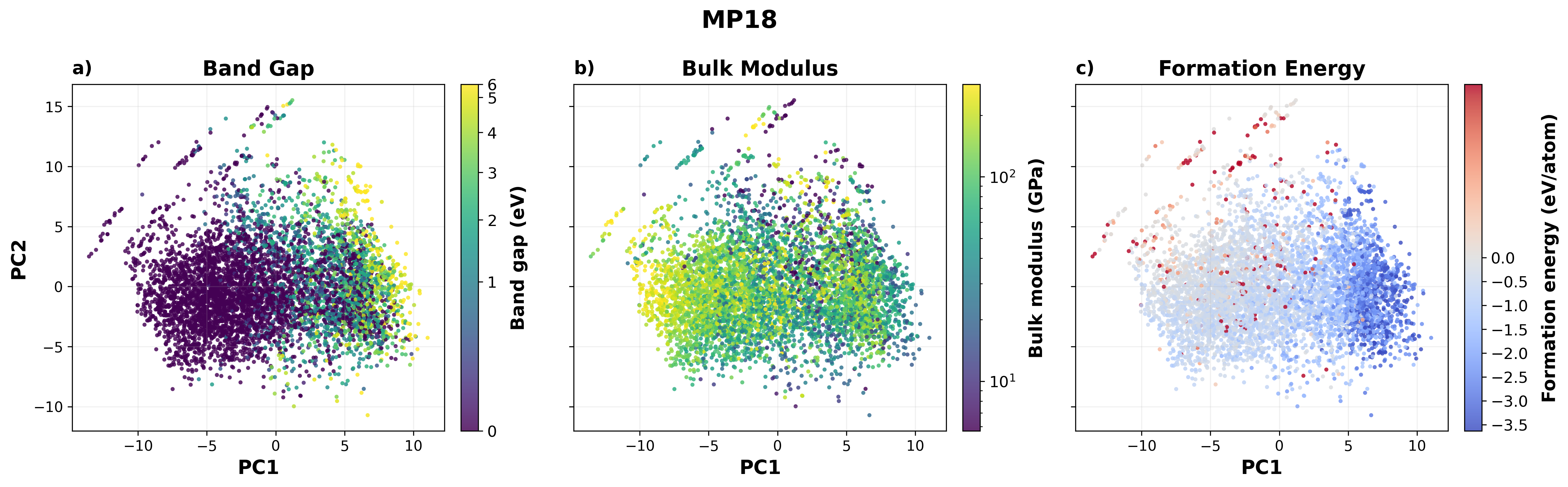}
\caption{
Properties distribution of MP18 shown in the first two principal components of feature space.
    }
\label{fig:DistributionOutcomesMP18}
\end{figure*}

\begin{figure*}[!ht]
        \centering
\includegraphics[width=\linewidth]{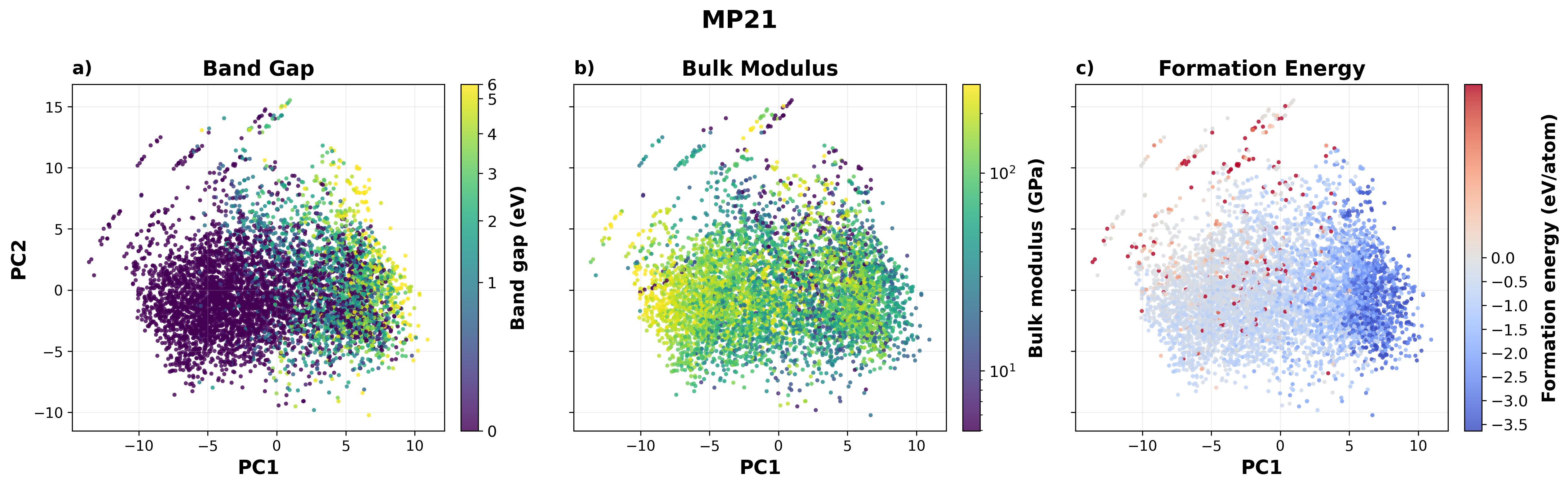}
\caption{ Properties distribution of MP21 shown in the first two principal components of feature space.  }
\label{fig:DistributionOutcomesMP21}
\end{figure*}

\begin{figure*}[!ht]
        \centering
\includegraphics[width=\linewidth]{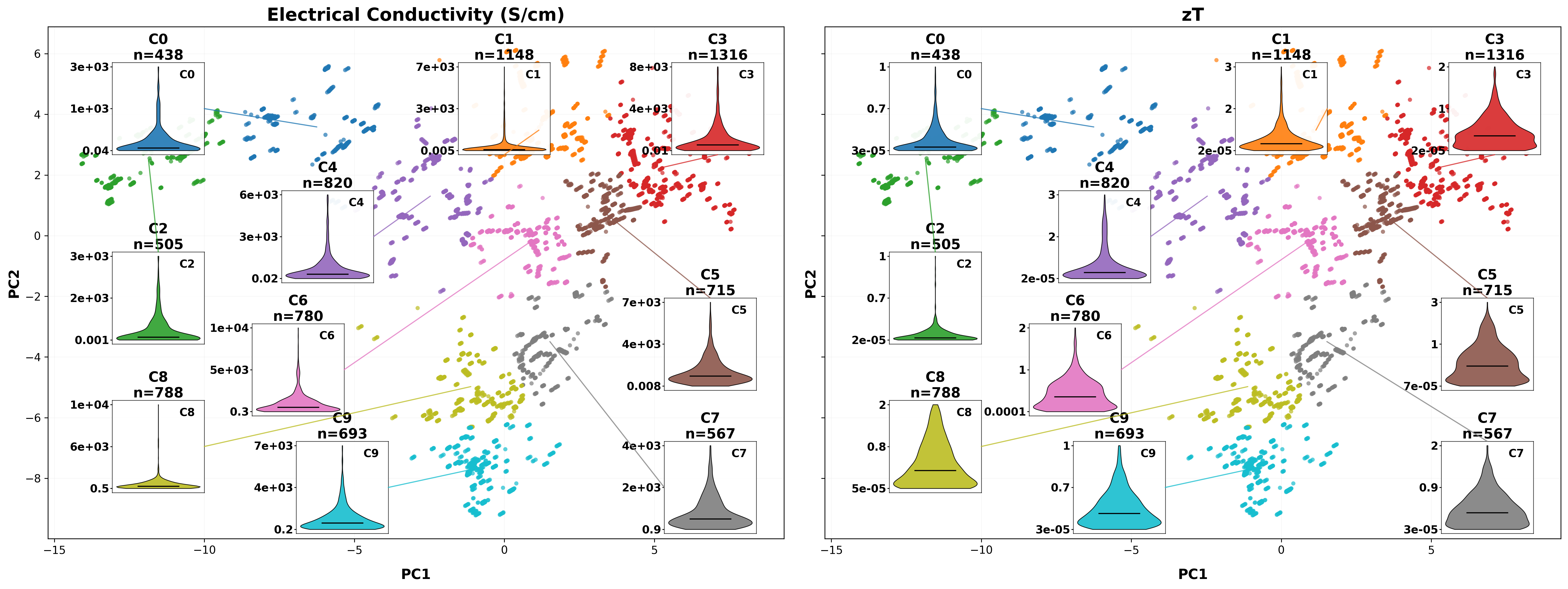}
\caption{
 Electrical conductivity and figure of merit (zT) distribution shown in the first two principal components of feature space in the sysTEm dataset.
    }
\label{fig:DistributionOutcomessysTEm}
\end{figure*}

\FloatBarrier
\section{Surrogate models hyperparameters}
\label{model_hyperparameters}

The surrogate models used in this work were implemented in Python using
\texttt{scikit-learn} and \texttt{xgboost}. The Random Forest models were
implemented with \texttt{RandomForestRegressor} from \texttt{scikit-learn}.
We used 100 trees, with one third of the input features considered at each
split. All available CPU cores were used during training. The random seed was
fixed to \texttt{random\_state = 0} to ensure reproducibility.

The XGBoost model was implemented with \texttt{XGBRegressor} from the
\texttt{xgboost} Python package. We used 1000 boosting estimators with a
learning rate of 0.25. L2 and L1 regularization were set to 0.01 and 0.1,
respectively. Feature subsampling was set to
$1/3$ per tree. We used four parallel trees per boosting
iteration and trained the model on CPU. The random seed was also fixed to
\texttt{random\_state = 0} for all XGBoost runs.
\end{document}